\documentclass{article}

\usepackage{arxiv}
\usepackage[utf8]{inputenc} 
\usepackage[T1]{fontenc}    
\usepackage{hyperref}       
\usepackage{url}            
\usepackage{booktabs}       
\usepackage{amsfonts}       
\usepackage{nicefrac}       
\usepackage{microtype}      
\usepackage{lipsum}
\usepackage{tabto}
\usepackage{graphicx}
\usepackage{caption}
\usepackage{float}
\usepackage{subfigure}
\usepackage{xcolor}
\usepackage{amssymb}
\usepackage{amsmath}
\usepackage{latexsym}
\usepackage{algorithm}
\usepackage{algpseudocode}
\usepackage[mathscr]{euscript}
\usepackage{url}
\usepackage{bm}
\usepackage{textcomp}
\usepackage{authblk}
\usepackage[superscript,biblabel]{cite}
\newcommand\norm[1]{\left\lVert#1\right\rVert}
\newcommand*\dif{\mathop{}\!\mathrm{d}}

\usepackage[title]{appendix}

\title{Reduced-Order Modeling of Unsteady Fluid Flow \\ Using Neural Network Ensembles}

\author[a,b]{Rakesh Halder \thanks{Corresponding Author: rhalder@umich.edu \\ \hspace*{1.5em} The presented research was conducted during the author's internship at Autodesk Research.}}
\author[b]{Mohammadmehdi Ataei}
\author[b]{Hesam Salehipour}
\author[a]{Krzysztof Fidkowski}
\author[c]{Kevin Maki}

\affil[a]{Department of Aerospace Engineering, University of Michigan, Ann Arbor, MI, USA 48109}
\affil[b]{Autodesk Research, Toronto, ON, CAN M5G 1M1}
\affil[c]{Department of Naval Architecture and Marine Engineering, University of Michigan, Ann Arbor, MI, USA 48109}

\date{\vspace{-5ex}}

\begin{document}
\maketitle

\begin{abstract}
The use of deep learning has become increasingly popular in reduced-order models (ROMs) to obtain low-dimensional representations of full-order models. Convolutional autoencoders (CAEs) are often used to this end as they are adept at handling data that are spatially distributed, including solutions to partial differential equations. When applied to unsteady physics problems, ROMs also require a model for time-series prediction of the low-dimensional latent variables. Long short-term memory (LSTM) networks, a type of recurrent neural network useful for modeling sequential data, are frequently employed in data-driven ROMs for autoregressive time-series prediction. When making predictions at unseen design points over long time horizons, error propagation is a frequently encountered issue, where errors made early on can compound over time and lead to large inaccuracies. In this work, we propose using bagging, a commonly used ensemble learning technique, to develop a fully data-driven ROM framework referred to as the CAE-eLSTM ROM that uses CAEs for spatial reconstruction of the full-order model and LSTM ensembles for time-series prediction. When applied to two unsteady fluid dynamics problems, our results show that the presented framework effectively reduces error propagation and leads to more accurate time-series prediction of latent variables at unseen points. 
\end{abstract}


\section{Introduction}

Physics-based numerical simulation has become an indispensable tool in engineering and scientific applications, allowing for accurate representations of complex physical phenomena. Physical simulations are formulated as sets of governing equations, typically in the form of parameterized partial differential equations (PDEs) that are discretized over a computational domain. Simulations often involve a set of design parameters $\bm{\mu}$ that govern aspects such as boundary conditions, the geometry of the computational domain, and physical properties. In tasks like design optimization, where numerous simulations are run for different designs, it is important to achieve high accuracy, or fidelity. However, high-fidelity simulation comes at a large computational cost, and this can lead to a bottleneck in the design optimization process.

Employing reduced-order models (ROMs) is a prevalent strategy to mitigate this high computational cost~\cite{lucia2004reduced}. By utilizing a limited set of training data from computed high-fidelity simulations, ROMs construct a low-dimensional surrogate model that can be evaluated in real-time and provide accurate full-order model (FOM) approximations at unseen designs within the parameter space covered by the training data. ROMs significantly reduce the degrees of freedom of the FOM through the use of a low-dimensional embedding that can be effectively mapped back to the full-order state, leading to a substantial decrease in computational cost. ROMs consist of two stages: an offline stage, which is computationally intensive, involving the computation of high-fidelity solutions by solving the FOM to generate data snapshots and train the low-dimensional surrogate model, and an online stage, where the model is used to approximate solutions at desired points. Running the FOM requires numerically solving the governing equations over the computational domain, which comes at a much larger computational cost than the ROM, which represents the FOM using a small number of variables. 

Most ROMs utilize the proper orthogonal decomposition~\cite{BerkoozStructure}, a linear method that uses the singular value decomposition (SVD) to obtain a low-rank subspace consisting of a number of linearly independent basis vectors. A linear combination of these basis vectors is computed using a set of expansion coefficients to approximate full-order states within the solution space. The POD basis vectors are interpretable, allowing for visualization of dominant physical features. Although POD is widely used, it encounters difficulties when applied to highly nonlinear problems, often requiring a large number of basis vectors to provide reasonable accuracy~\cite{HESTHAVEN201855}.

More recently, machine learning and artificial intelligence (AI) methods have been used to provide nonlinear mappings between the low-dimensional embedding and high-fidelity solution space. In particular, deep learning~\cite{lecun2015deeplearning} approaches have been used to develop ROMs that provide efficient nonlinear manifolds of physical systems. Convolutional autoencoders (CAEs), a type of artificial neural network, have been used to build ROMs and shown to outperform POD-based methods ~\cite{Lee2020ModelRO,halder2022non}. Artificial neural networks that use convolutional layers efficiently learn structures and patterns present in spatially distributed data, including the solutions to PDEs. Autoencoders consist of two individual neural networks: an encoder, which maps high-dimensional inputs to a low-dimensional latent space, and a decoder, which maps the low-dimensional latent space to a reconstruction of the high-dimensional input. Although autoencoders can effectively learn nonlinear relationships, they lack interpretability. Recent works have used variational autoencoders~\cite{eivazi2022towards,solerarico2023betavariational} to incorporate interpretable nonlinear modes into ROMs, although this results in lower reconstruction accuracy when compared to vanilla autoencoders. Another approach for simulating systems governed by PDEs at a reduced cost is the use of neural operators~\cite{wang2021learning,li2021physics}, which implicitly learn the governing equations using deep neural networks. However, this approach typically requires very large amounts of training data to cover parameter spaces effectively, limiting their use in design optimization.

ROMs are categorized as either non-intrusive or projection-based; non-intrusive ROMs use a fully data-driven approach, while projection-based ROMs incorporate the governing equations to solve a low-dimensional version of the FOM. While projection-based ROMs can provide better accuracy and more physically consistent results, they incur a larger computational cost and often exhibit stability issues~\cite{GRIMBERG2020109681} that inhibit convergence. Additionally, they are generally not portable between different solvers. 

ROMs are developed for both steady and unsteady physics problems. The first approach involves computing a single point estimate of the low-dimensional embedding, while the latter involves making predictions over a prescribed time horizon. Unsteady non-intrusive ROMs combine either POD or CAE for spatial reconstruction of full-order states with a model used to make time-series predictions of the low-dimensional embeddings. Deep learning methods are popular for this as well, including long short-term memory (LSTM) neural networks~\cite{rahman2019nonintrusive} and transformer neural networks~\cite{solerarico2023betavariational}. Both LSTMs and transformers are powerful models for handling data that are sequential in nature such as time-series data. Transformers utilize an architecture that is much more complex than the one LSTMs use, which can lead to better performance for complex time-series problems and other tasks such as developing large language models~\cite{zhao2023survey}. However, this comes at an increased cost for both training and inference when compared to LSTMs, which use a significantly lower number of parameters. The use of LSTMs for time-series forecasting is well-established~\cite{gers2000learning,hill1996neural}, making them a popular choice.

When making time-series predictions at unseen data sets over a long time horizon, error propagation is a common issue. Errors made in early predictions can accumulate and compound over time, leading to substantial inaccuracies. Unseen data sets are particularly suspect to this, as the model may not account for shifting data patterns. Additionally, there is no feedback from previously seen data to correct errors. This phenomenon is commonly observed in data-driven computational fluid dynamics (CFD) applications~\cite{lee2019data,vinuesa2022enhancing}. As a result, most studies in the literature focus on applying unsteady non-intrusive ROMs to single-parameter problems~\cite{rahman2019nonintrusive, solerarico2023betavariational, wu2022non}, where the ROM is both trained on and used for a single design. We have identified two previously published studies that combine CAEs and LSTMs for non-intrusive ROMs and apply them to unseen designs~\cite{maulik2021reduced, hasegawa2020machine}. The work by Maulik et al. does not mention the error propagation issue, while the work by Hasegawa et al. presents results of low accuracy. A recent paper~\cite{jeon2024residual} by Jeon et al. introduced a hybrid AI-CFD method using flow residuals to address the issue of error propagation. However, this method is intrusive and requires the ROM to have access to the CFD solver for computing the residuals, which leads to a more computationally intensive online stage.

In this work, we propose the use of ensemble learning~\cite{sagi2018ensemble}, a machine learning technique for improving the stability and lowering the variance of predictive models. Ensemble learning involves combining multiple base models, referred to as \emph{weak learners}, to create a composite model that offers greater accuracy. To this end, we employ bootstrap aggregating (bagging) as the ensemble learning method for the temporal portion of the ROM, which involves training the weak learners on subsets of the dataset chosen randomly through sampling with replacement. The fully data-driven framework is referred to as the CAE-eLSTM ROM, and our results show that using ensembles leads to significantly improved stability and predictive performance when applied to two unsteady, incompressible, laminar fluid dynamics problems using different CFD solvers.

\section{Methods}
In this section, we give an overview of both CAEs and LSTMs and how they are combined to develop an unsteady ROM using bagging as an ensemble learning method. CAEs are used to both provide a low-dimensional latent representation of the solution space and provide spatial reconstructions of full-order states. A temporal forecasting model of the latent variables is also required when using unsteady ROMs. LSTMs, a type of recurrent neural network, are used in this work. Although we do not provide direct comparisons to POD-based ROMs, a brief introduction to POD is given in this section. Comparisons of reconstruction errors using POD and CAE are given in Appendix~\ref{appendix:pod}.

\subsection{Proper Orthogonal Decomposition}
Using training data obtained from a set of $n$ solution snapshots calculated at chosen design points in the parameter space, a snapshot matrix, $\bm{X} \in \mathbb{R}^{N\times n}$, can be assembled.
\begin{equation}
\bm{X} \in \mathbb{R}^{N\times n}= [{\bm{x}}^1, {\bm{x}}^2, \cdots , {\bm{x}}^n] = [{\bm{x}}(\bm{\mu}^{1}), {\bm{x}}(\bm{\mu}^{2}), \cdots , {\bm{x}}(\bm{\mu}^{n})].
\end{equation}
A subspace $\bm{\mathcal{V}}$ associated with $\bm{X}$ exists such that $\bm{\mathcal{V}}$ = span$(\bm{X})$. It assumed that $\bm{\mathcal{V}}$ provides a good approximation of the solutions for $ \bm{\mu}$ within the parameter space if there is a sufficient variety of training data in $\bm{X}$. A rank $k$ set of orthonormal basis vectors $[{\bm{\psi}}^{1}, {\bm{\psi}}^{2}, \cdots, {\bm{\psi}}^{k}] \in \mathbb{R}^{N}$,  where $k \ll N$, is associated with $\bm{\mathcal{V}}$ such that each solution $\bm{x}^{i}$ in $\bm{X}$ can be reconstructed as a linear combination of the basis vectors 

\begin{equation}
\bm{x}^{i} \approx a^{i}_{1}\bm{\psi}^{1} + a^{i}_{2}\bm{\psi}^{2} \cdots + a^{i}_{k}\bm{\psi}^{k},
\end{equation}

where $\bm{a}^{i}$ are the low-dimensional expansion coefficients. 
The truncated singular value decomposition (SVD) of $\bm{X}$ decomposes the matrix into  two orthonormal matrices $\bm{U} \in \mathbb{R}^{N\times n}$ and $\bm{V} \in \mathbb{R}^{n\times n}$, and a diagonal matrix $\bm{\Sigma} \in \mathbb{R}^{n\times n}$ such that
\begin{equation}
\bm{X = U\Sigma V^{T}}.
\end{equation}
$\bm{U}$ contains a set of $n$ left singular vectors that span the column space of $\bm{X}$, $\bm{V}$ contains a set of $n$ right singular vectors that span the row space of $\bm{X}$, and diag($\bm{\Sigma}$) $\in \mathbb{R}^{n}= [{\sigma}_1, {\sigma}_2, \cdots, {\sigma}_n]$ contains the singular values in decreasing order, $\sigma_{1} \geq \cdots \geq \sigma_{n} \geq 0$. The first $k$ vectors of $\bm{U}$ form the POD basis, $\bm{\Psi} \in \mathbb{R}^{N\times k} = [{\bm{\psi}}^{1}, {\bm{\psi}}^{2}, \cdots, {\bm{\psi}}^{k}]$. The singular values are a measure of the amount of information represented by each vector. Often, the singular values decay rapidly, and only the first $k$ singular vectors are chosen to form the POD basis to preserve the most important features of the solution space. The relative information content $E$ of the subspace is  used in practice to choose a value of $k$,
\begin{equation}
    E(k) = \dfrac{\sum_{j=1}^{k}\sigma_{j}^{2}}{\sum_{j=1}^{n}\sigma_{j}^{2}}, 
    \label{eqn:sigma}
\end{equation}
and $k$ is chosen such that $E(k) \geq \gamma$, where $\gamma \in $ [0,1] is usually set to a value $\gamma \geq 0.95$ \cite{MrosekROM}. Approximations of full-order solutions at unseen design parameters ${\bm{x}}(\bm{\mu}^{*})$ are obtained as
\begin{equation}
\bm{x}(\bm{\mu}^{*}) \approx \bm{\Psi}\bm{a}^{*} = a^{*}_{1}\bm{\psi}^{1} + a^{*}_{2}\bm{\psi}^{2} \cdots + a^{*}_{k}\bm{\psi}^{k}.
\end{equation}

\subsection{Convolutional Autoencoders}

Convolutional neural networks were initially developed for computer vision applications, where they have been shown to vastly outperform traditional statistical image processing techniques~\cite{lecun1989handwritten, russakovsky2015imagenet}. They have also demonstrated high predictive performance when used in ROMs~\cite{xu2020multi, Lee2020ModelRO}, highlighting their versatility. Autoencoders are a type of feedforward neural network that aim to accurately reconstruct inputs in the output layer, $g: \bm{x} \rightarrow \bm{\tilde{x}}$ where $\bm{x} \approx \bm{\tilde{x}}$. Autoencoders are composed of two individual feedforward neural networks. The encoder $g_{\text{enc}}: \mathbb{R}^{N} \rightarrow \mathbb{R}^{k}$ where $k \ll N$ maps a high-dimensional input $\bm{x}$ into the low-dimensional latent space $\bm{a}$, and the decoder $g_{\text{dec}}: \mathbb{R}^{k} \rightarrow \mathbb{R}^{N}$ maps the latent variables back to an approximation of the high-dimensional input  $\bm{\tilde{x}}$. The combination of the two results in
\begin{equation}
    g: \bm{\tilde{x}} = g_{\text{dec}} \circ g_{\text{enc}}(\bm{x}).
\end{equation}
Deep neural networks provide nonlinear function maps through the use of activation functions. During training, the neural network parameters are tuned using backpropagation~\cite{Rumelhart:1986we}, an algorithm utilizing automatic differentiation that aims to minimize a differentiable loss function that measures the discrepancy between the computed and true outputs. This allows for discovering highly complex and abstract functional relationships that do not follow any strong model assumptions. The number of trainable parameters in vanilla artificial neural networks grows large with the number of hidden layers and neurons, and can lead to a very large computational training cost.

\begin{figure}[!h]
\centering
\includegraphics[width=0.95\textwidth]{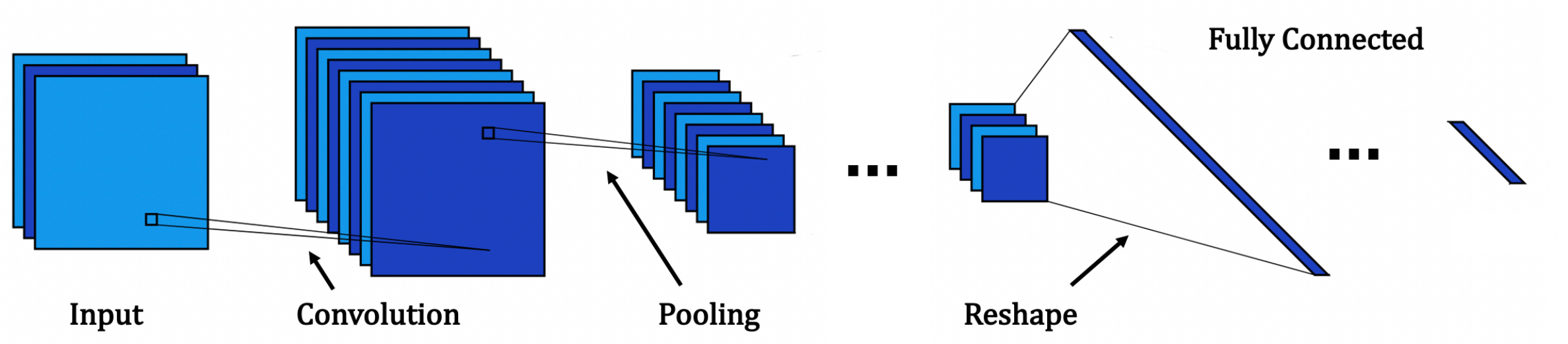}
\caption{Architecture of the encoder of a convolutional autoencoder (CAE) consisting of convolutional, pooling, and fully connected layers.}
\label{fig:cae_encoder}
\end{figure}

Convolutional neural networks effectively implement parameter sharing to limit the total number of trainable parameters in the network, where rather than weight combinations existing for each pair of neurons between layers, multiple neurons share a single weight. Convolutional layers use a number of filters that convolve over the input data, with each filter having its own set of weights. Pooling layers are also used in convolutional networks to summarize the features in input layers through operations including averaging and maximization. The input layer to a convolutional neural network is composed of a number of channels, with each representing a different state. In images, this is generally the levels of red, green, and blue present in each pixel, while for physical simulation data the states can represent components of the normalized velocity, pressure, density, etc. Once an autoencoder is sufficiently trained and $g(\bm{x}) \approx \bm{x}$ for all inputs over the training dataset $\bm{X}$, the corresponding latent variables $\bm{a}$ can be passed to the decoder $g_{\text{dec}}(\bm{a})$ to obtain accurate approximations $\bm{\tilde{x}}$ for all data in $\bm{X}$. High-dimensional data existing outside of the training set $\bm{x}^{*}$ may also be well-approximated if an accurate approximation of the latent variables $\bm{a}^{*}$ is obtained. Figure~\ref{fig:cae_encoder} shows an example of the encoder section of a convolutional autoencoder, which is composed of convolutional, pooling, and fully connected layers. A similar architecture in reverse would form the decoder. 

The input and output layers of CAEs are often structured as two-dimensional states within each channel when used for two-dimensional physical simulations. To accomodate this structure, training data needs to be reshaped before being input into the network through the use of a reshape operator $\bm{R}$,
\begin{equation}
    \bm{R}: \mathbb{R}^{N \times n_{c}} \rightarrow \mathbb{R}^{n_{y} \times n_{x} \times n_{c}},
\end{equation}
where $n_{y}$ refers to the number of data points in the vertical direction and $n_{x}$ the number of data points in the horizontal direction. The reshape operator is applied to each separate state that occupies the $n_{c}$ input channels.

\subsection{Long Short-Term memory (LSTM) Neural Networks}

\begin{figure}[!h]
\centering
\includegraphics[width=0.6\textwidth]{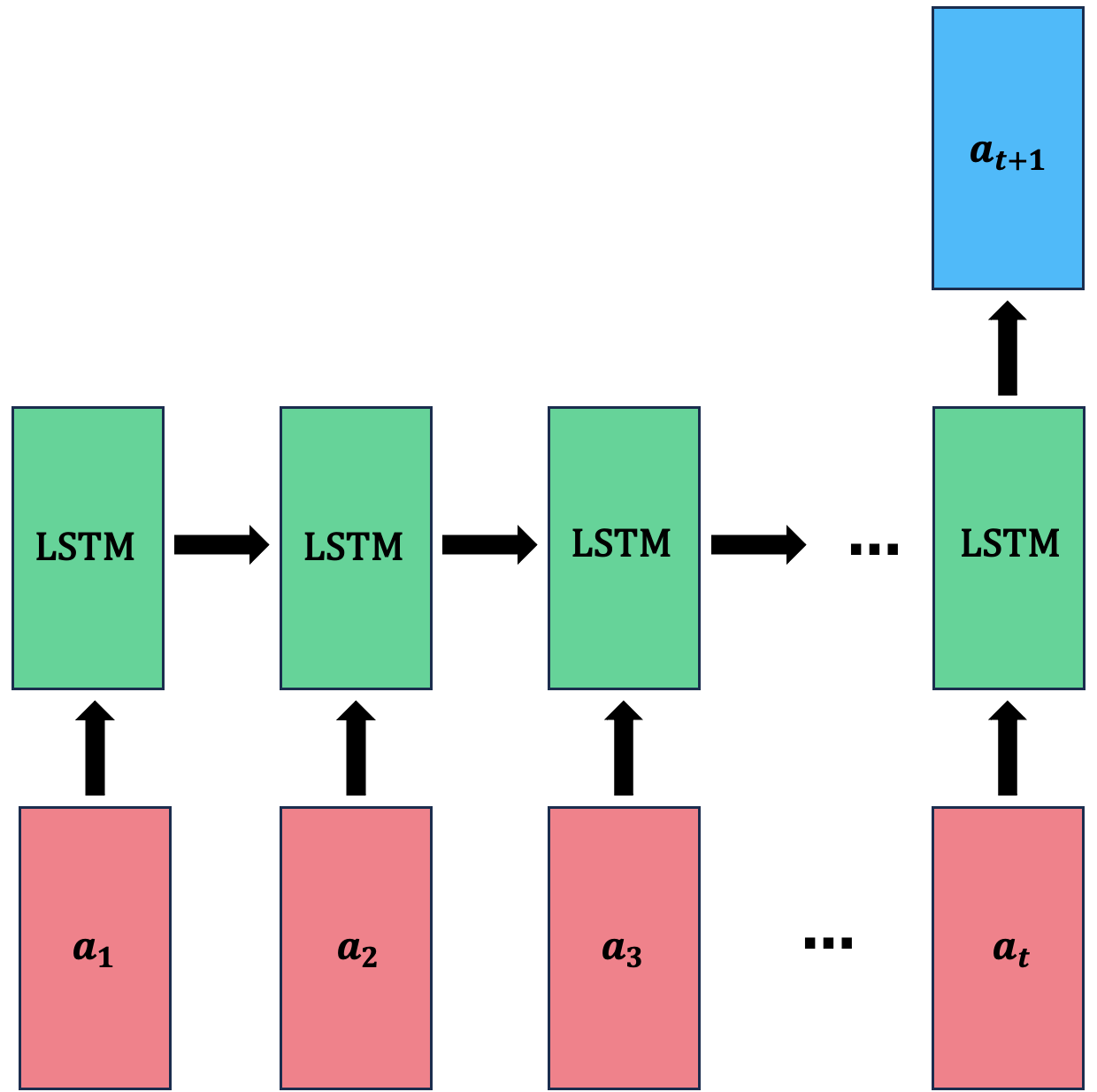}
\caption{Diagram of a single-layer LSTM neural network making a prediction one timestep ahead.}
\label{fig:lstm}
\end{figure}

Traditional recurrent neural networks are well-suited for handling sequential data; however, they have difficulties learning and capturing long-range dependencies. A primary reason for this is the vanishing gradient problem~\cite{hochreiter1998vanishing}, which occurs when gradients shrink significantly as they are back-propagated through time, which can cause the model to \emph{forget} early sequence information. LSTMs were introduced~\cite{hochreiter1997long} to address this issue by using a more complex network architecture that includes a series of gates which effectively control the flow of information. An individual LSTM cell consists of a cell state $\bm{c}_{t}$ which acts as an internal memory, and a hidden state $\bm{h}_{t}$,  which serves as an output. The cell state is responsible for selectively retaining or forgetting information from previous steps, while the hidden state represents the model's cumulative output. LSTM cells contain a forget gate $\bm{f}_{t}$, input gate $\bm{i}_{t}$, and output gate $\bm{o}_{t}$. Forget gates determine what information from previous states should be forgotten. Input gates decide what new pieces of information should be stored in the current state. Output gates use the current and previous cell states to determine what information is retained in the model. Using this architecture, the model is able to retain information that is relevant for long-range sequential dependencies and discard information that becomes irrelevant over time, making LSTMs a popular choice for sequential modeling.  For a given input $\bm{a}_{t} \in \mathbb{R}^{k}$, the LSTM equations are given as

\begin{equation}
\begin{aligned}
\bm{f}_{t} = \sigma (\mathcal{F}_{f}(\bm{a}_{t})) \\
\bm{i}_{t} = \sigma (\mathcal{F}_{i}(\bm{a}_{t})) \\
\bm{o}_{t} = \sigma (\mathcal{F}_{o}(\bm{a}_{t})) \\
\bm{\tilde{c}}_{t} = \text{tanh}(\mathcal{F}_{c}(\bm{a}_{t})) \\
\bm{c}_{t} = \bm{f}_{t} \odot \bm{c}_{t-1} + \bm{i}_{t} \odot \bm{\tilde{c}}_{t} \\
\bm{h}_{t} = \bm{o}_{t} \odot \text{tanh}(\bm{c}_{t}),
\end{aligned}
\end{equation}

where $\sigma$ is the sigmoid function, tanh is the hyperbolic tangent function, and $\odot$ is the Hadamard product. $\mathcal{F}$ is a linear function of the weights $\bm{W}_{a} \in \mathbb{R}^{n_{h}}$ and $\bm{W}_{h} \in \mathbb{R}^{n_{h}}$ and biases $\bm{b} \in \mathbb{R}^{n_{h}}$, where $n_{h}$ is the number of neurons in the hidden layer of each LSTM cell, and is given as

\begin{equation}
\mathcal{F} = \bm{W}_{a}\bm{a}_{t} + \bm{W}_{h}\bm{h}_{t-1} + \bm{b}.
\end{equation}
A unique set of weights and biases belongs to each gate or cell state. When using LSTMs for time-series prediction, an autoregressive prediction is used, often referred to as the sliding window approach. For a given input sequence containing multivariate data for $w$ timesteps, the next step in the sequence is predicted. When making the next prediction, the current prediction is incorporated into the input sequence by removing the first element and shifting the rest to the prior position. Eventually, the model inputs will consist of only previously computed predictions, which makes model performance sensitive to error propagation. The input sequence length, or window size $w$, is an important hyperparameter to consider when training LSTMs. Although LSTMs are designed to effectively learn long-range dependencies, using an input sequence length that is too long can lead to the inclusion of outdated and irrelevant information. On the other hand, using an input sequence length that is too small can ignore important long-range information. The optimal choice is highly problem-dependent; similar problems found in the literature can be consulted, or methods such as cross-validation can be used to find an optimal value. A diagram of a single-layer LSTM making a prediction one timestep ahead is shown in Figure~\ref{fig:lstm}.

\subsection{Ensemble Learning}

While LSTMs are powerful models for time-series prediction, the quality of predictions over long time horizons is highly sensitive to the weights and biases of the model. Additionally, the total number of weights and biases is usually very large, which makes finding an optimal configuration very difficult, even when using state of the art optimization algorithms.

To mitigate the issue of high model variance, ensemble learning is a commonly used approach. By leveraging multiple base models that exhibit high variance, ensemble methods significantly reduce errors and improve robustness. Boosting and bootstrap aggregating (bagging) are the two main types of ensemble learning methods. Boosting~\cite{freund1999short} trains individual models sequentially, where each subsequent weak learner focuses on correcting prediction errors from the previous ones. While boosting is widely used and offers good performance, the base models must be trained iteratively, which incurs a large computational cost, especially in the context of neural networks.

\begin{figure}[!h]
\centering
\includegraphics[width=0.95\textwidth]{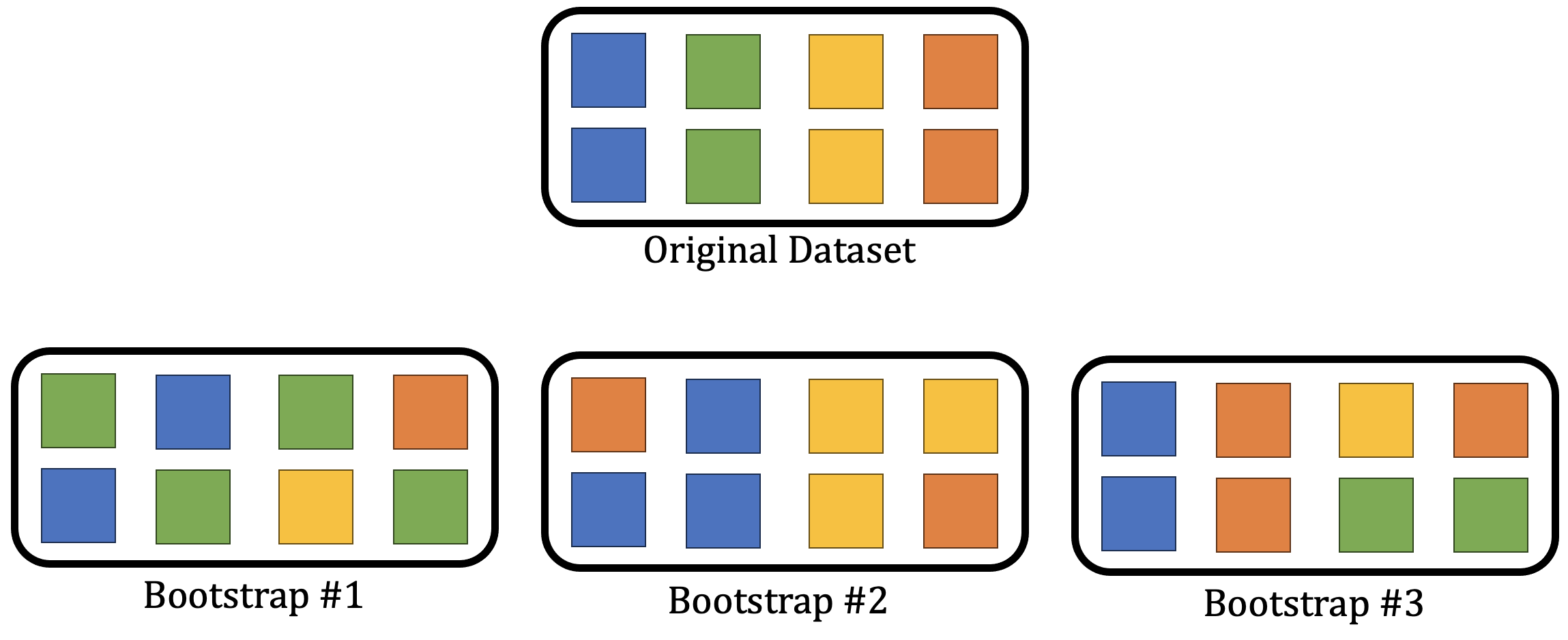}
\caption{An example of bootstrapping, where random subsets of the original dataset are chosen through sampling with replacement.}
\label{fig:bagged}
\end{figure}

Bagging~\cite{breiman1996bagging} is an algorithm that consists of two stages: bootstrapping and aggregation. Bootstrapping is a resampling technique where multiple random subsets of the data set, chosen through sampling with replacement, are constructed. The subsets of the data typically contain the same number of data points as the original data set. This approach leads to individual data points being present multiple times in the individual subsets. An example of bootstrapping is shown in Figure~\ref{fig:bagged}, where the bootstrapped data sets contain more or less instances of the original data points. Multiple base models are trained individually on each of the bootstrapped data sets, which can be done in parallel. The aggregation stage creates an ensemble model by taking an average of the predictions given by each weak learner. Given $m$ weak learners $f_{i}$, the aggregate prediction $\bar{f}$ is given as

\begin{equation}
  \bar{f} = \frac{1}{m}\sum_{i}^{m} f_{i}.
\end{equation}

Bagging reduces the issue of error propagation by averaging the errors of multiple weak learners, which will become increasingly small as the number of learners grows, leading to much lower variance. Bagging also offers a considerable gain in robustness and generalization by leveraging the diversity of the weak learners to increase the capability of handling varying patterns outside of the training data. As an increasing number of weak learners are added, the reduction in variance plateaus, leading to diminishing returns in performance gains. After a certain point, the computational cost of adding more weak learners outweighs the marginal improvement in predictive performance. As the complexity of the problem increases, the number of weak learners required to make accurate predictions also generally increases. While theoretical considerations can be made to analyze the trade-offs between bias, variance, and covariance to determine the optimal number of weak learners, this number is difficult to choose a priori. A reasonable choice is based on both the problem's complexity and similar problems found in the literature, which are not yet present in our case and are established in this work. 

 \subsection{CAE-eLSTM ROM}

This section describes the non-intrusive ROM framework combining convolutional autoencoders for spatial reconstruction and LSTM ensembles utilizing bagging for temporal prediction, which we will refer to as a CAE-eLSTM ROM. The offline stage is computationally intensive, requiring multiple solves of the FOM in addition to training multiple neural networks. First, solutions to the FOM for designs $\bm{\mu}$ in the training set of design parameters $\bm{\mathcal{U}}_{\text{train}} \in \mathbb{R}^{n\times p}$ are obtained for all timesteps and are assembled into a matrix $\bm{X}$. Next, a convolutional autoencoder with latent dimension $k$ is trained on $\bm{X}$ for a sufficient number of epochs such that inputs are accurately reconstructed. Then, the expansion coefficients for the training data $\bm{A}_{\text{train}}$ are computed by feeding the full-order states from $\bm{X}$ through the encoder $g_{\text{enc}}$. Sequences of length $w$ are then generated from $\bm{A}_{\text{train}}$ and $m$ LSTMs using the same set of initial weights and biases are trained on individual data subsets chosen randomly through sampling with replacement. PyTorch~\cite{paszke2019pytorch}, an open source deep learning framework available for Python, is used to implement the autoencoder and LSTM using default weight and bias initialization settings.

\begin{figure}[!h]
  \centering
  \includegraphics[width=0.95\textwidth]{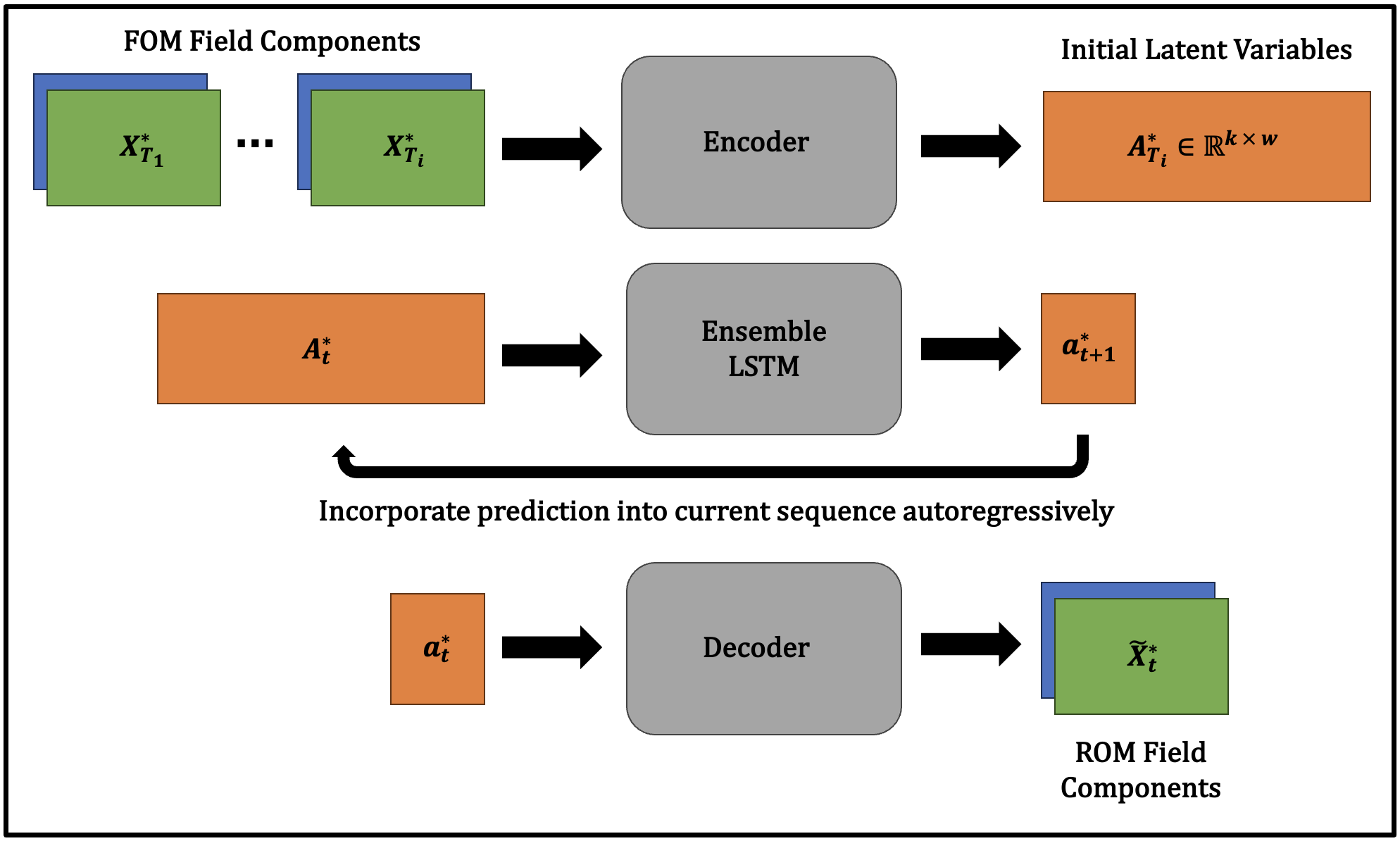}
  \caption{Schematic of the online stage of the CAE-eLSTM ROM.}
  \label{fig:diagram}
  \end{figure}

A schematic describing the online stage of the ROM is shown in Figure~\ref{fig:diagram}. The online stage involves executing the ROM to compute approximate solutions at a point $\bm{\mu}^{*}$. First, the FOM is run for $T_{i}$ timesteps, as an initial sequence of latent variables is required for the LSTM ensemble. The computed full-order states are fed through the encoder to obtain the initial sequence of latent variables $\bm{A}^{*}_{T_{i}} \in \mathbb{R}^{k \times w}$. The FOM must be run for at least $T_i \geq w$ timesteps, and potentially longer depending on the how useful simulation data from the initial timesteps are for the ROM. The latent variables for the rest of the prediction horizon are calculated autoregressively using the LSTM ensemble by taking an average of the predictions from the individual weak learners. After the predicted latent variables are found for each timestep $t$, they are fed through the decoder to obtain approximate full-order fields. The online and offline stages are outlined in Algorithm~\ref{alg:rom}. 

When using a ROM with a single LSTM as in previous works~\cite{maulik2021reduced,hasegawa2020machine}, the LSTM is trained on the entire time-series dataset and is used alone for time-series prediction. When compared to these works, the main difference in our presented ROM framework is the use of an ensemble model for time-series prediction. Although this results in an increased offline training cost, using an ensemble allows for better accuracy and stability over long prediction horizons. While the use of an ensemble LSTM is simple, it can greatly improve ROM performance without requiring additional training data.

\begin{algorithm}[!htbp]
\caption{Offline and online stages of CAE-eLSTM ROM}\label{alg:rom}
\begin{algorithmic}[1]
\Function{CAE\_ eLSTM\_ OFFLINE}{$\bm{\mathcal{U}}_{\text{train}}, k, m, w$}
\State Compute high-fidelity solutions for $\bm{\mu} \in \bm{\mathcal{U}_{\text{train}}} $ by solving FOM and assemble into $\bm{X}$.
\State Train convolutional autoencoder with latent dimension $k$ on $\bm{X}$.
\State Calculate expansion coefficients for training data $\bm{A}_{\text{train}} = g_{\text{enc}}\left(\bm{X} \right)$.
\State Train $m$ LSTMs $\mathcal{E} = [\text{LSTM}_{1}(\bm{A}), \text{LSTM}_{2}(\bm{A}), \cdots \text{LSTM}_{m}(\bm{A})]$ using a window size of $w$ on sequences sampled randomly with replacement from $\bm{A}_{\text{train}}$.
\State \Return $\left(g_{\text{enc}}, g_{\text{dec}}, \mathcal{E} \right)$
\EndFunction
\end{algorithmic}
\[\]
\begin{algorithmic}[1]
\Function{CAE\_ eLSTM\_ ONLINE}{$\bm{\mu}^{*},g_{\text{enc}}, g_{\text{dec}}, \mathcal{E}$}
\State Run FOM for $T_{i}$ timesteps at $\bm{\mu}^{*}$ and compute the expansion coefficients $\bm{A}^{*}_{T_{i}} = g_{\text{enc}}\left([\text{X}^{*}_{T_1}, \text{X}^{*}_{T_2}, \cdots \text{X}^{*}_{T_i}] \right)$.
\For{$t \in \{T_{i}, T_{i+1},\dots,T-1\}$}
\State Compute $\bm{a}^{*}_{t+1} = \mathcal{E}(\bm{A}^{*}_{t})$ by using an average of the $m$ individual LSTM predictions.
\State Incorporate $\bm{a}^{*}_{t+1}$ into the current window $\bm{A}^{*}_{t}$.
\EndFor
\State Predict full-order fields $\bm{\tilde{X}}^{*} = g_{\text{dec}}\left(\bm{{A}}^{*}\right)$
\State \Return $\bm{\tilde{X}}^{*}$
\EndFunction
\end{algorithmic}
\end{algorithm}

\section{Results}

The two test cases used to demonstrate the performance of the CAE-eLSTM ROM are a lid-driven cavity and the flow over a cylinder. Both cases use two-dimensional computational domains. The lid-driven cavity case consists of three design parameters controlling the geometry of the domain, while the cylinder case consists of two design parameters controlling geometric and physical properties of the simulation. Different CFD solvers are used for each case; this is done to illustrate the portability of the ROM and allow for the use of a structured grid for the cylinder case. Latin hypercube sampling (LHS)~\cite{lhs}, a popular statistical method for generating samples from a multi-dimensional parameter space is used to generate sets of design parameters that are split into training and test sets. LHS aims to maximize the distance and minimize the correlation amongst samples produced by using uniform sampling. The ROM is used to predict the vertical and horizontal components of the velocity. The training data for each velocity component is scaled to a range [0,1] using min-max scaling before input into the CAE. Similarly, the CAE latent variables are also scaled using min-max scaling before being used for the LSTM. Data normalization improves performance and allows the optimizer to learn optimal network parameters at a much faster rate~\cite{Sola1997ImportanceOI}. Both the CAE and LSTM are trained using the mean squared error loss function. An NVIDIA DGX system, consisting of eight NVIDIA A100 GPUs, is used to train the CAE and LSTMs, run the cylinder simulations, and for ROM inference. The lid-driven cavity simulations are run on a local workstation using a single CPU core. 

The performance of the ensemble ROM is tested against a CAE-LSTM ROM that uses a single LSTM model. Using five different random initializations of the weights and biases prescribed by different seeds, the time-series of latent variables are visually compared in addition to performance metrics averaged over the seeds. When training multiple LSTMs on bootstrapped data sets for a single seed $s$, the same initial weights and biases are used for each individual weak learner. Evaluating the ROM's performance using different seeds allows for testing its sensitivity to the initial weights and biases. $N_s = 5$ different seeds are used to initialize the LSTM model, with $s = $[1, 2, 3, 4, 5]. For a single test point $\bm{\mu}^{*}$ and seed $s$, an error metric of interest is the relative $L^{2}$ error in a field component averaged over the prediction time horizon,

\begin{equation}
        \epsilon_{s} =  \frac{1}{T-T{i}} \sum_{t=T_{i}+1}^{T} \dfrac{\norm{\bm{x}^{t} - \tilde{\bm{x}}^{t}}^{2}} {\norm{\bm{x}^{t}}^2}.
\end{equation}

The errors are then averaged over each seed to give a single error metric $\bar{\epsilon}$

\begin{equation}
        \bar{\epsilon} =   \frac{1}{N_s} \sum_{s=1}^{N_s} \epsilon_{s}.
\end{equation}

We are also interested in how sensitive the error is to the initial weights and biases of the model. The sample standard deviation $\sigma_{s}$ of $\epsilon_{s}$ is used as a measure  of this,

\begin{equation}
        \sigma_{s} = \text{std}(\epsilon_{s}).
\end{equation}

A lower standard deviation in the error term indicates that the model performance over the prediction horizon does not vary much with the initial weights and biases.

\subsection{Lid-Driven Cavity}

The first test case is a geometrically parameterized two-dimensional lid-driven cavity flow, a popular benchmark problem for CFD solvers. Unsteady, incompressible, laminar flow is simulated by solving the  Navier-Stokes equations, given as

\begin{equation}
  \label{eqn_mass}
  \int_S {\overrightarrow{V}} \cdot \dif \vec{n} \, dS =0,
  \end{equation}
  \begin{equation}
    \label{eqn_momentum}
 \int_\Omega {\frac{\partial \overrightarrow{V}}{\partial t} \dif \Omega} + \int_S  {\overrightarrow{V}} {\overrightarrow{V}} \cdot \dif \vec{n} \, dS  + \int_\Omega \nabla p  \dif \Omega - \int_S \nu (\nabla {\overrightarrow{V}}+\nabla {\overrightarrow{V}}^T) \cdot \dif \vec{n} \, dS  = 0,
  \end{equation}

\begin{figure}[!h]
\centering
\includegraphics[width=0.49\textwidth]{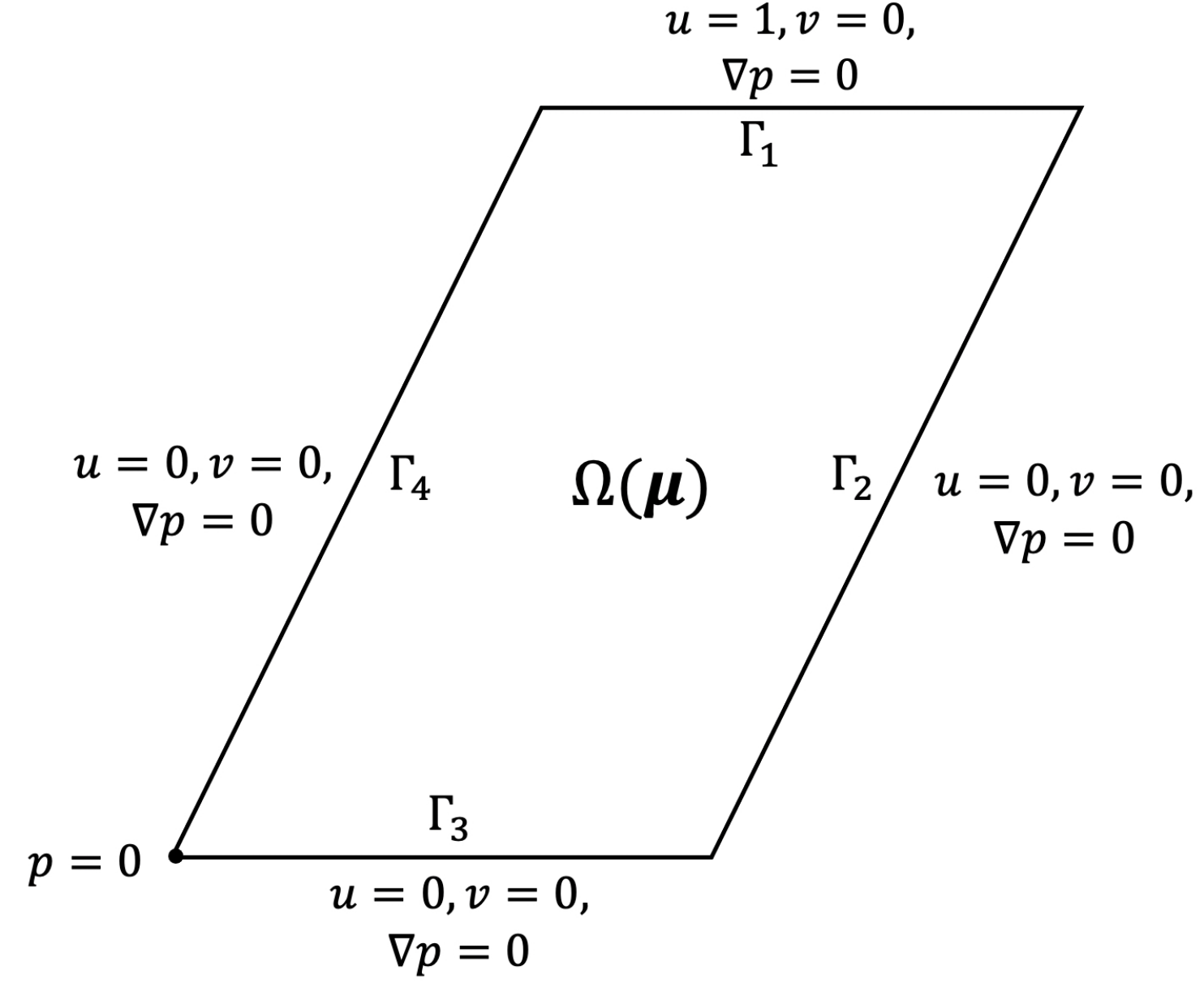}
\includegraphics[width=0.49\textwidth]{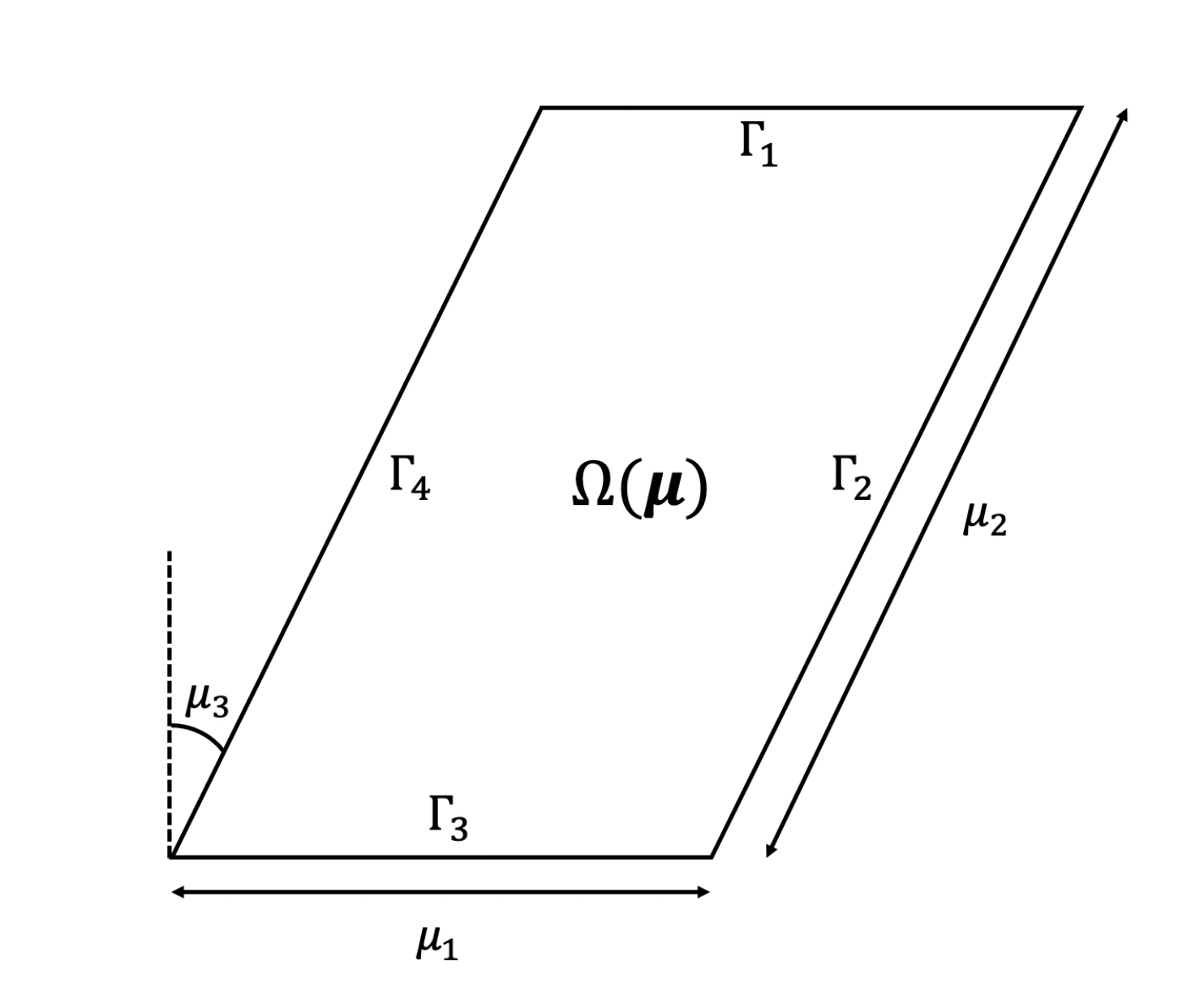}
\caption{Schematics describing the lid-driven cavity problem.}
\label{fig:cavity}
\end{figure}

where ${\overrightarrow{V}}$ = [$u$, $v$] is the velocity vector, $\Omega$ is the fluid domain, $S$ is the face-area vector, $\vec{n}$ is the outward-pointing normal, $\nu$ is the kinematic viscosity, and $p$ is the pressure. OpenFOAM~\cite{weller1998tensorial}, an open-source toolbox for multiphysics simulation is used. The boundary conditions and design parameters are shown in Figure~\ref{fig:cavity}. On each edge $\Gamma_{i}, i \in [1,2,3,4]$ of the computational domain $\Omega$, the pressure gradient $\nabla p$ is set to 0. No slip conditions exist on all of the edges except the top, where $u = 1, v = 0$. The reference pressure is set to $p = 0$ at the bottom left corner. The first design parameter $\mu_{1}$ controls the horizontal length, the second $\mu_{2}$ controls the slanting length, while the third $\mu_{3}$ controls the slanting angle. The Reynolds number $Re$ is set to 400, and is related to the kinematic viscosity as

\begin{equation}
  Re = 400 = \frac{\text{max}(\mu_{1}, \mu_{2})}{\nu(\bm{\mu})}.
\end{equation}

The bounds for the design parameters are given as
\begin{align*}
& \mu_{1} \in [1.2, 1.8], \\
& \mu_{2} \in [1.2, 1.8], \\ 
& \mu_{3} \in [-\frac{\pi}{6}, \frac{\pi}{6}].
\end{align*}

\begin{figure}[!h]
  \centering
  \subfigure[$\bm{\mu} = (1.5, \frac{3}{\sqrt{3}}, -\frac{\pi}{6})$]{\includegraphics[scale=0.23]{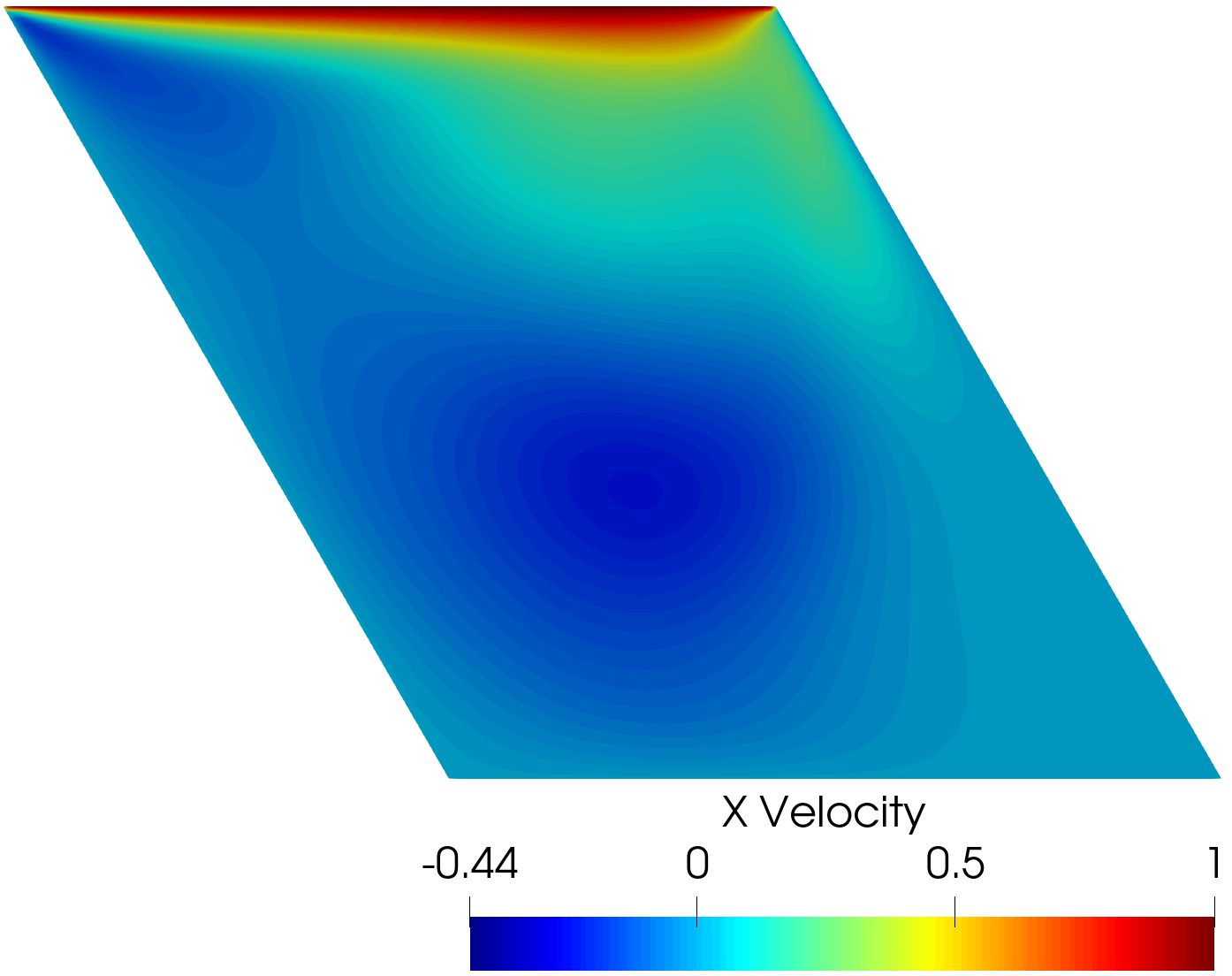}}
  \hspace{0.2cm}
  \subfigure[$\bm{\mu} = (1.5, 1.5, 0)$]{\includegraphics[scale=0.23]{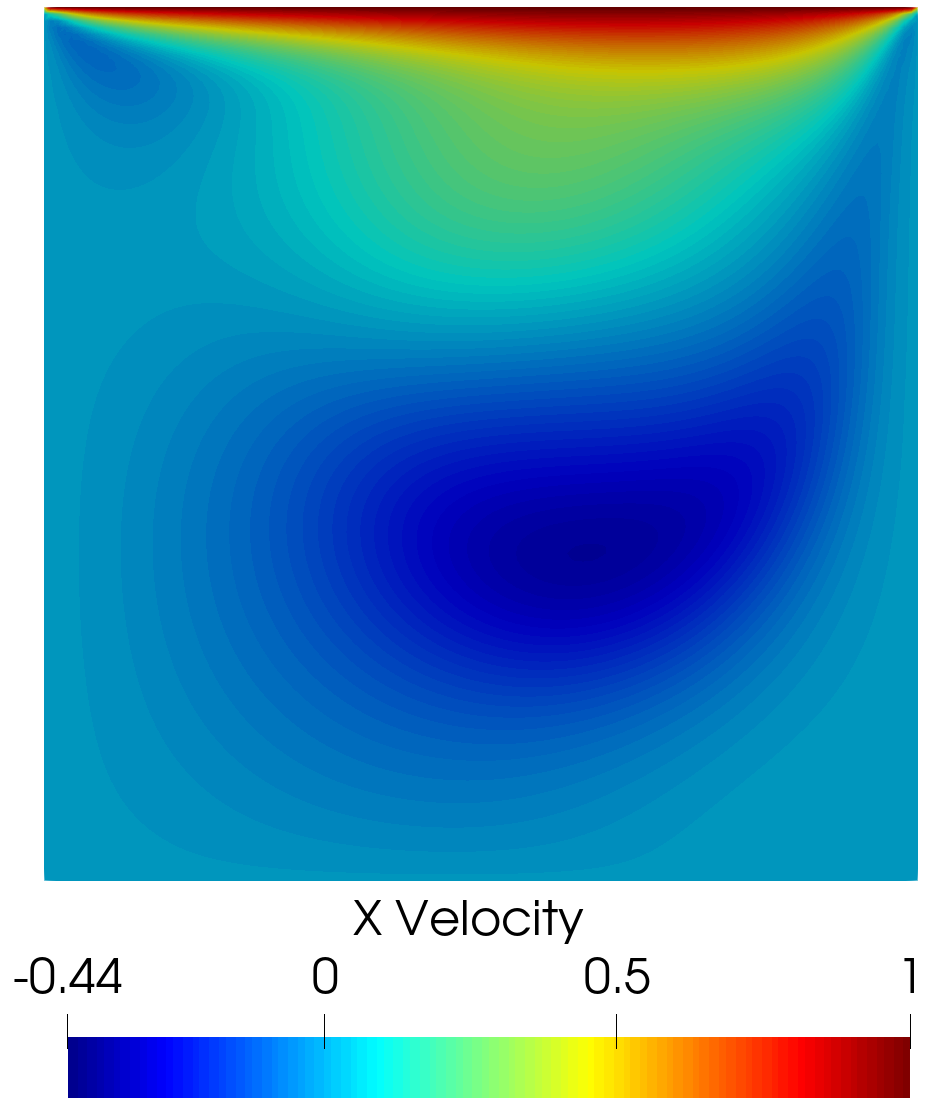}} 
  \hspace{0.2cm}
  \subfigure[$\bm{\mu} = (1.5, \frac{3}{\sqrt{3}}, \frac{\pi}{6})$]{\includegraphics[scale=0.23]{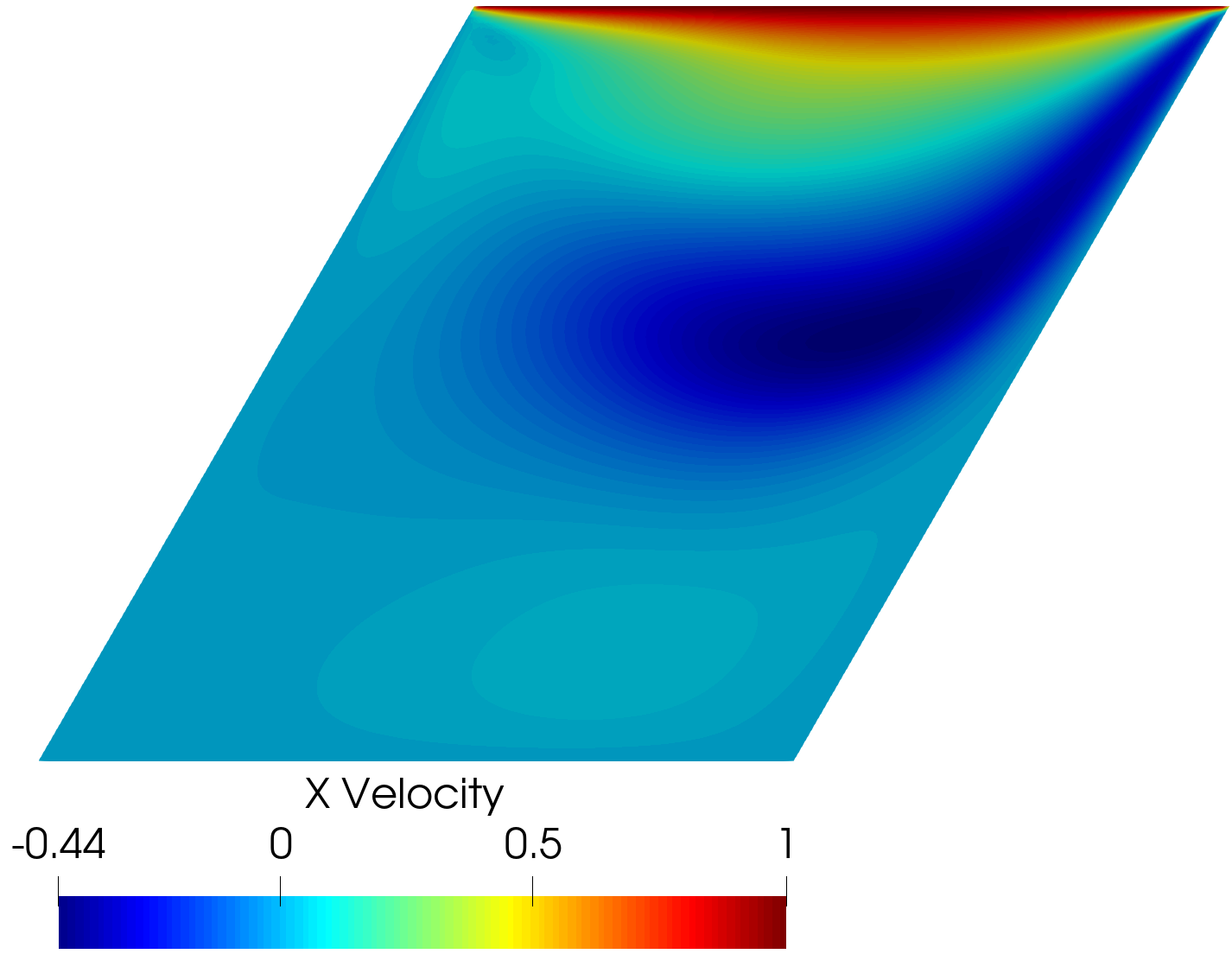}}
  \subfigure[$\bm{\mu} = (1.5, \frac{3}{\sqrt{3}}, -\frac{\pi}{6})$]{\includegraphics[scale=0.23]{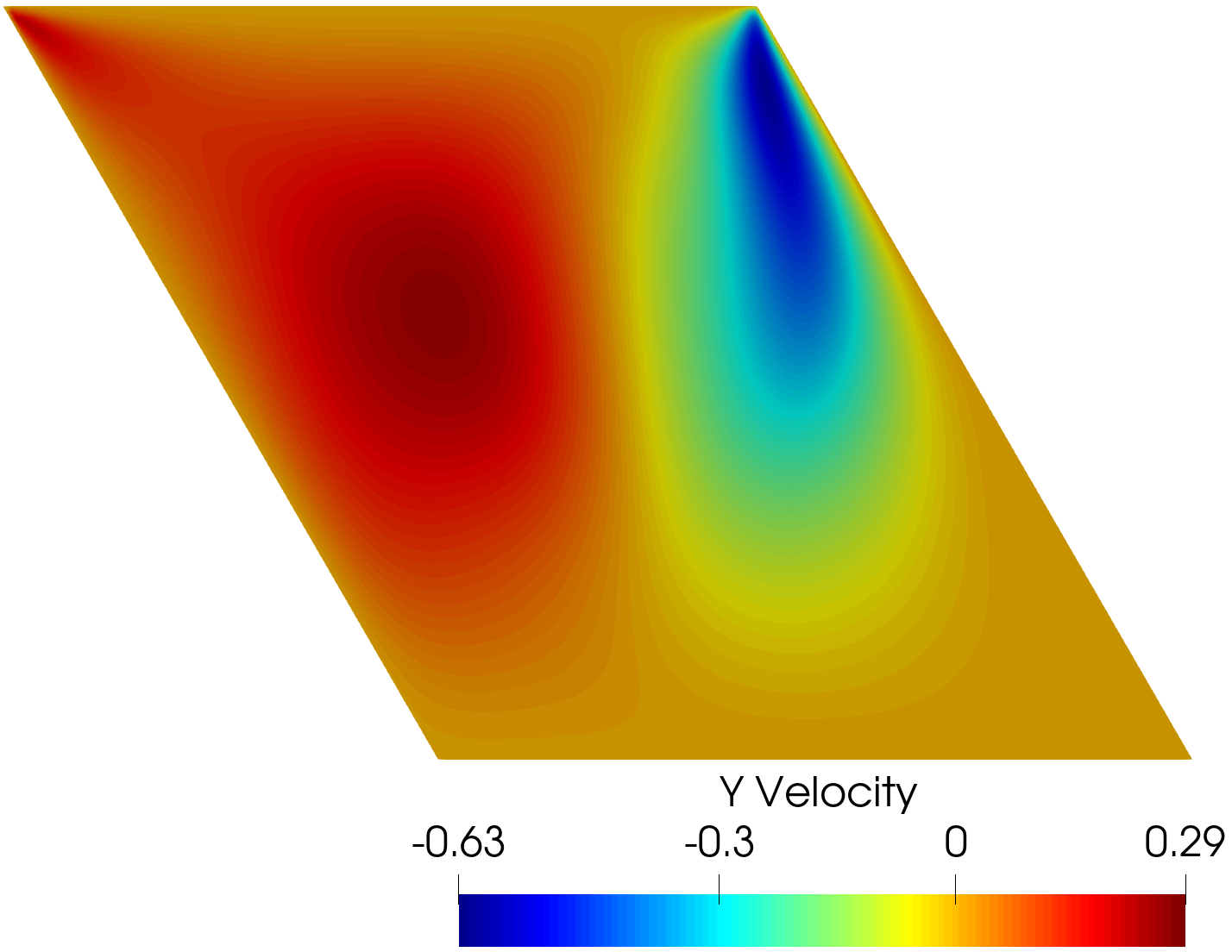}} 
  \hspace{0.2cm}
  \subfigure[$\bm{\mu} = (1.5, 1.5, 0)$]{\includegraphics[scale=0.23]{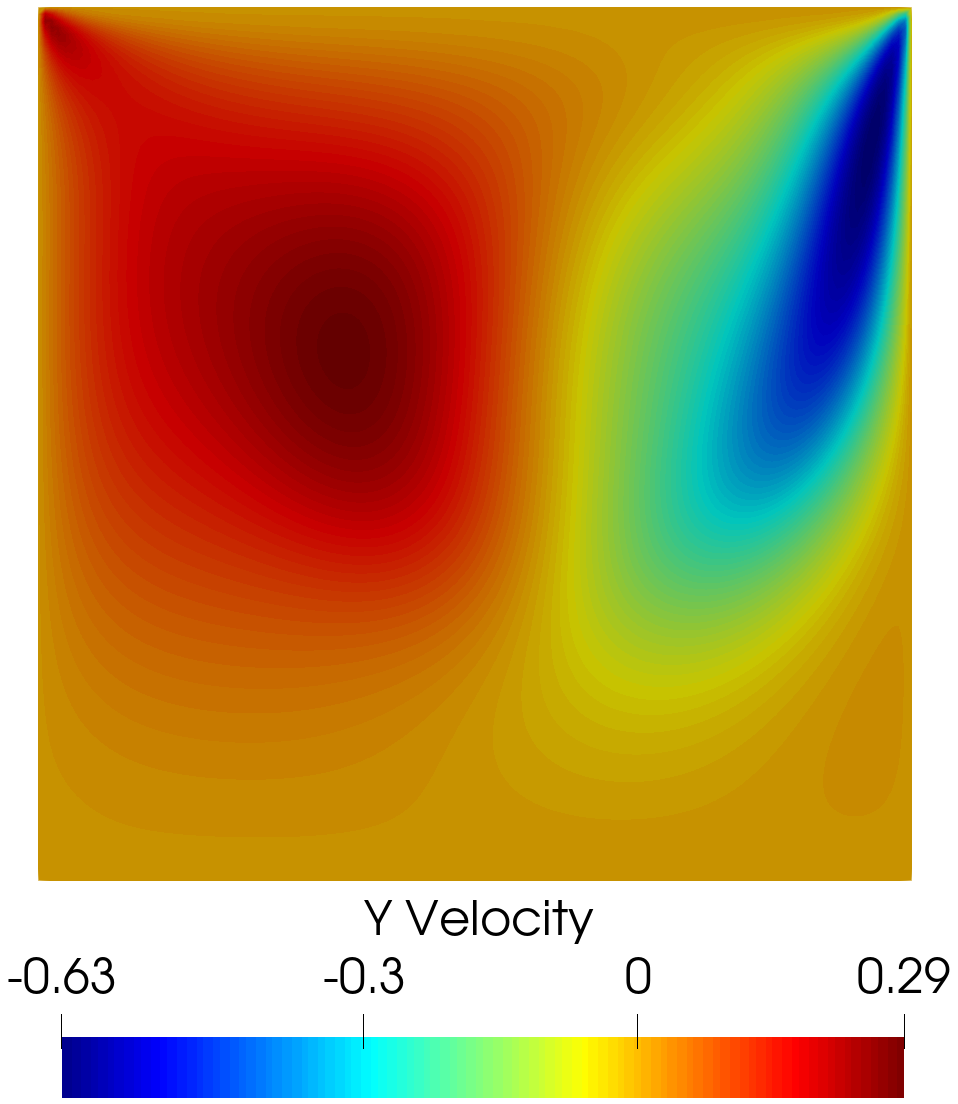}} 
  \hspace{0.2cm}
  \subfigure[$\bm{\mu} = (1.5, \frac{3}{\sqrt{3}}, \frac{\pi}{6})$]{\includegraphics[scale=0.23]{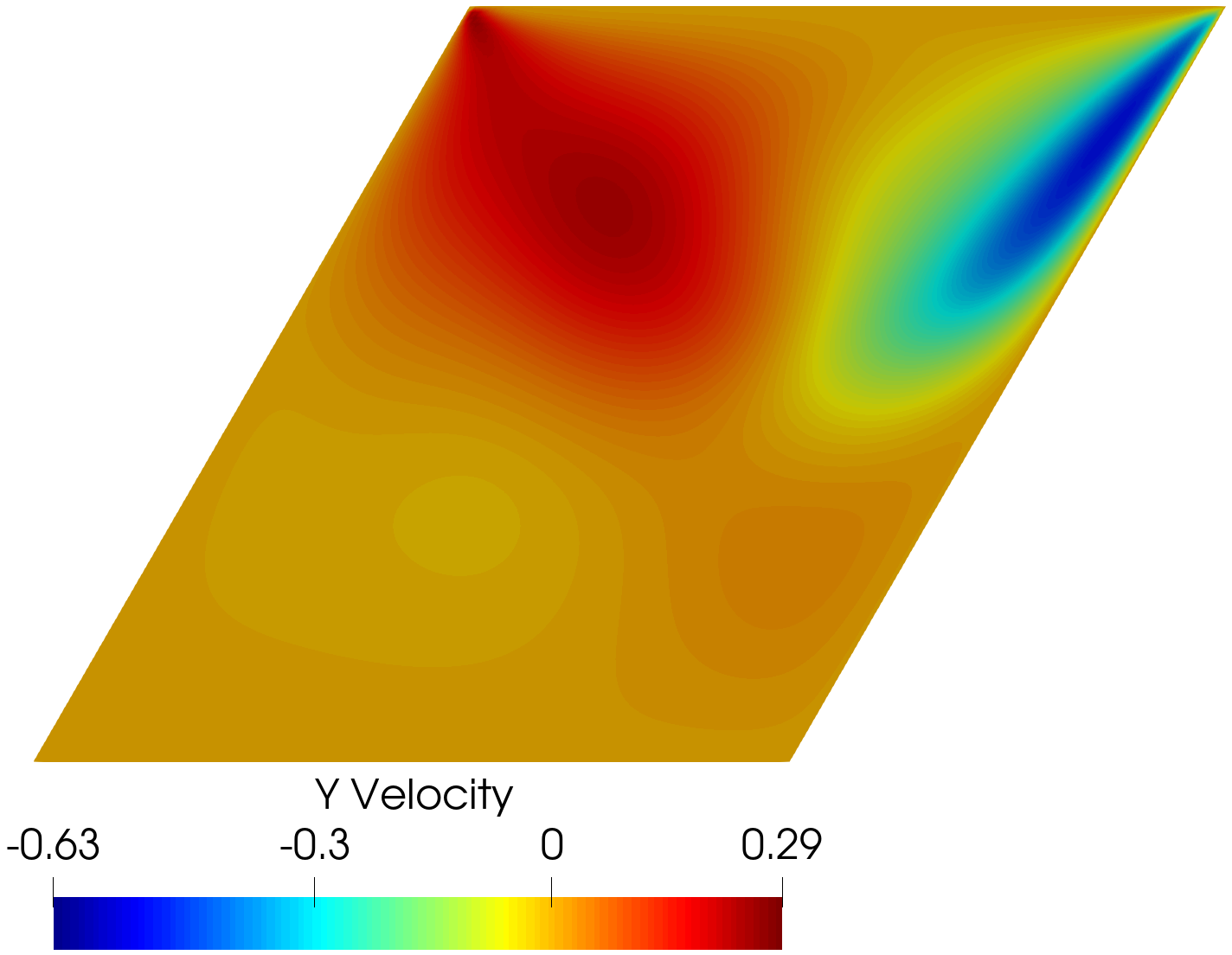}}
  \caption{Snapshots at $T = 5$ of $u$ (top) and $v$ (bottom) for the lid-driven cavity problem at three different sets of design parameters.}
  \label{fig:cavity_contours}
  \end{figure}

The computational mesh consists of $128 \times 128$ cells distributed uniformly in the $x$ and $y$ directions. A single cell exists in the spanwise direction $z$, resulting in $N = 16384$.  The initial condition is the solution to steady, incompressible, laminar flow at $Re = 20$ which is computed using the standard OpenFOAM solver simpleFoam. Unsteady flow is simulated using the OpenFOAM solver icoFoam for $T = 5$ seconds with data being saved every 0.025 seconds, leading to 200 time snapshots for a single simulation. Both solvers utilize the finite volume method (FVM). Figure~\ref{fig:cavity_contours} shows contours of $u$ and $v$ at $T = 5$ at three different sets of design parameters. A vortex that varies in shape, size, and location with the design parameters is shown for both velocity components. The variation of the flow with $\bm{\mu}$ is highly nonlinear, making this a difficult prediction problem for ROMs.

100 sets of design parameters are generated and randomly split into 90 training samples and 10 test points, which are given in Table~\ref{tab:cavity_dv}. The CAE-eLSTM ROM uses $m$ = 64 bagged LSTMs and a window size $w = 20$. These parameters were chosen through a trial-and-error process with the goal of maximizing predictive accuracy while trying to keep the computational cost of training the LSTM ensemble relatively low. This was done by directly evaluating the performance on the test set, while in practice methods like cross-validation would be used. Initial guesses for the window size were based on similar problems found in the literature. An initial number of approximately 50 weak learners was used, which was increased in increments of 8 until the performance gains plateaued. The CAE latent dimension is set to $k = 4$; below this value, the CAE reconstruction errors were found to be worse, and increasing $k$ offered no improvement. The LSTM architecture consists of two hidden layers consisting of 50 neurons each and a dropout~\cite{srivastava2014dropout} rate of 0.1. The output layer contains a sigmoid activation function so the outputs are constrained to a range of $[0,1]$. The CAE architecture is given in Appendix~\ref{appendix:arch}. The Adam optimizer~\cite{Adam} is used to train both the CAE and LSTM, with an initial learning rate of $\eta = 5 \times 10^{-4}$ and weight decay of $\lambda = 1 \times 10^{-6}$. Dropout and weight decay are used to prevent overfitting. The CAE is trained for a total of 200 epochs, while an individual LSTM is trained for 250 epochs. At the test points, the FOM is simulated for $T_{i} = 0.75$ seconds (30 snapshots), or 15\% of the total time horizon. The final 20 of these snapshots are used as the initial input sequence for the LSTM ensemble.

  \begin{table}[!h]
    \centering
    \begin{tabular}{@{}llll@{}}
    \toprule
    Test Case Index & $\mu_{1}$ & $\mu_{2}$ & $\mu_{3}$ \\ \midrule
    1               & 1.473     & 1.701     & 0.288     \\ 
    2               & 1.659     & 1.653     & -0.372    \\ 
    3               & 1.593     & 1.611     & 0.455     \\ 
    4               & 1.209     & 1.443     & 0.371     \\ 
    5               & 1.299     & 1.689     & -0.0367   \\ 
    6               & 1.371     & 1.311     & -0.351    \\ 
    7               & 1.443     & 1.617     & -0.487    \\ 
    8               & 1.232     & 1.335     & -0.330    \\ 
    9               & 1.719     & 1.785     & -0.455    \\ 
    10              & 1.281     & 1.449     & -0.340    \\ \bottomrule
    \end{tabular}
    \caption{Test case design parameters for the lid-driven cavity problem.}
    \label{tab:cavity_dv}
    \end{table}

\begin{figure}[!h]
  \centering
  \includegraphics[width=0.42\textwidth]{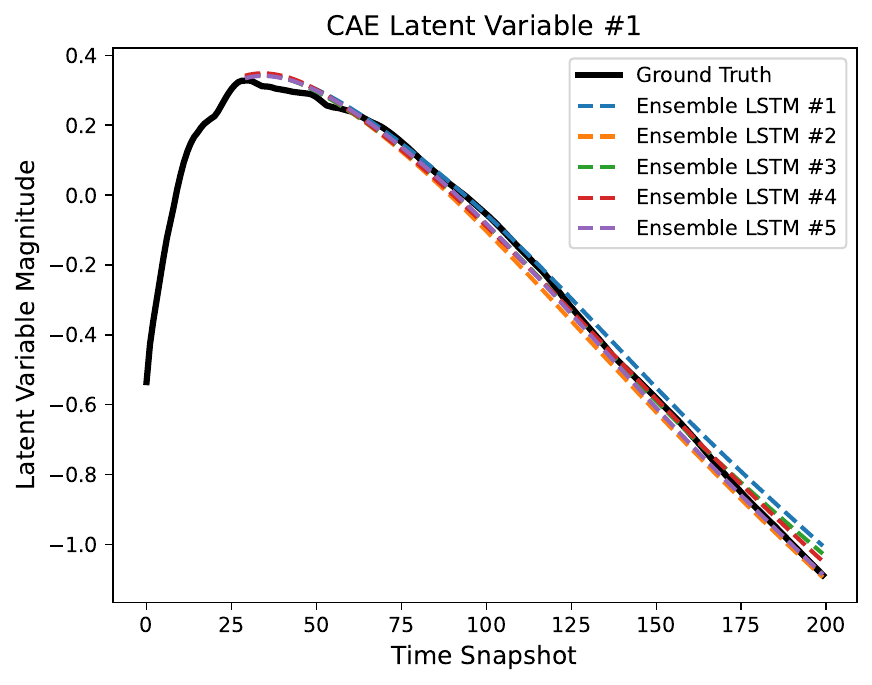}
  \includegraphics[width=0.42\textwidth]{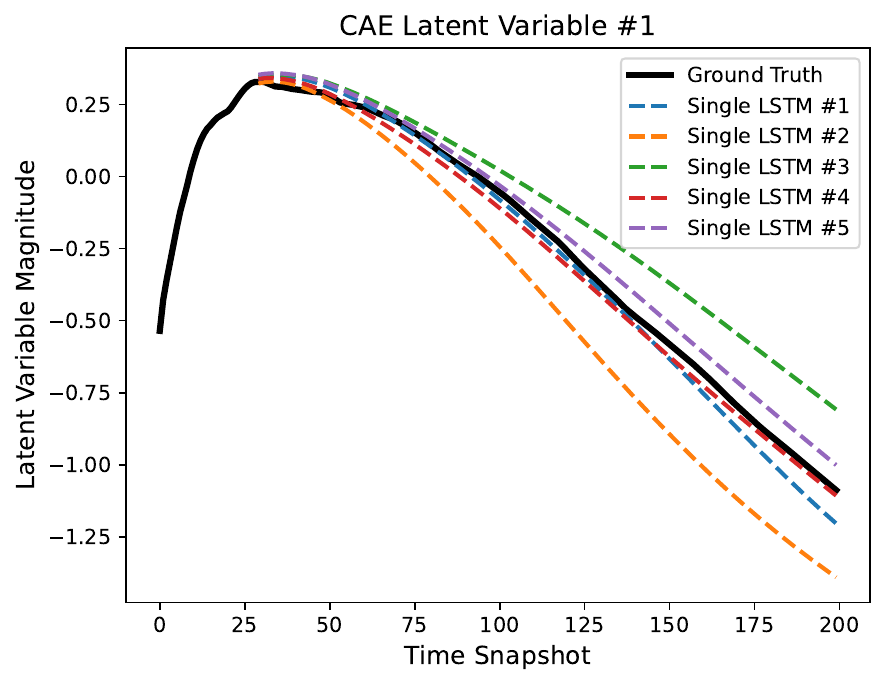}
  \includegraphics[width=0.42\textwidth]{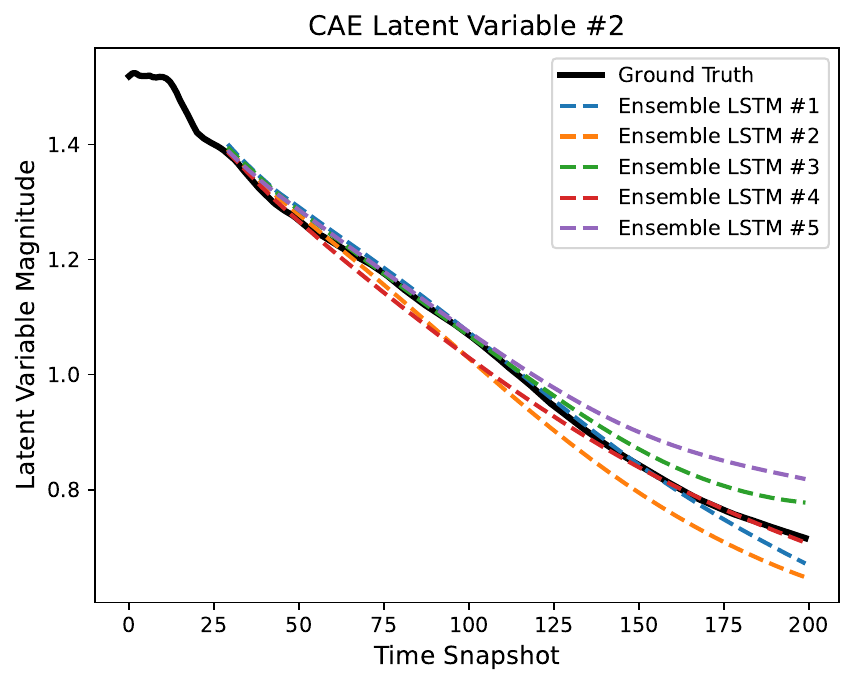}
  \includegraphics[width=0.42\textwidth]{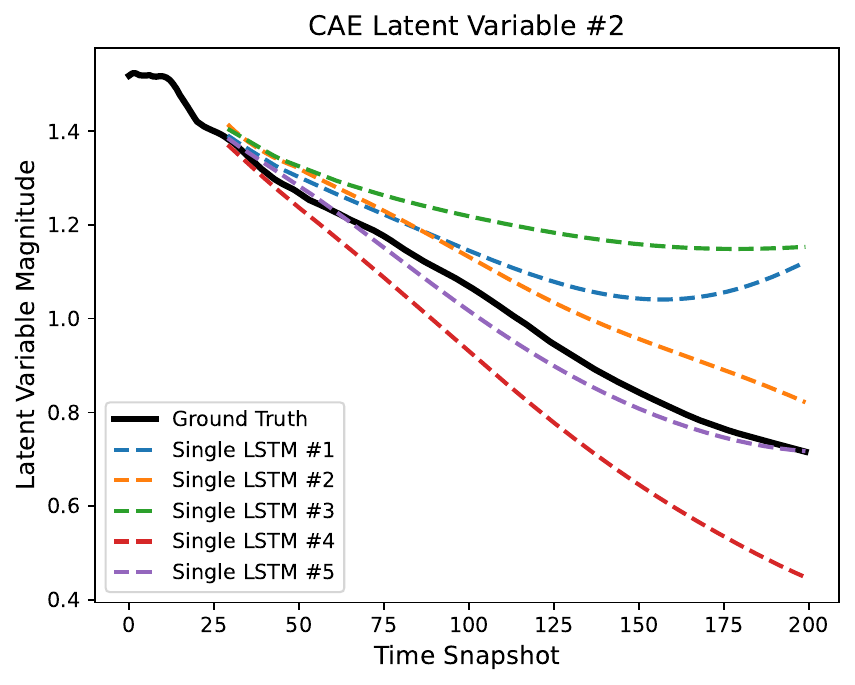}
  \includegraphics[width=0.42\textwidth]{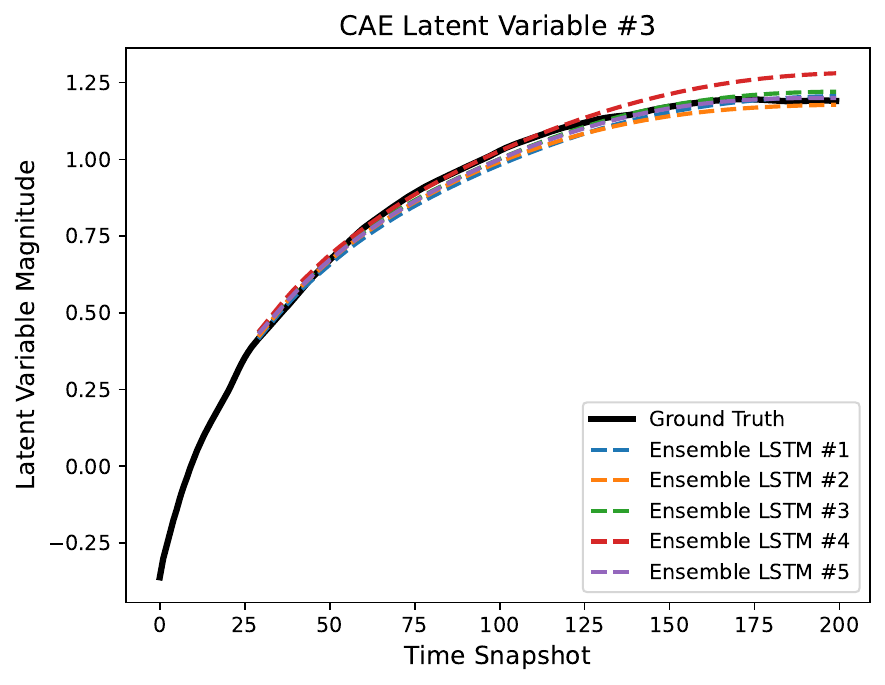}
  \includegraphics[width=0.42\textwidth]{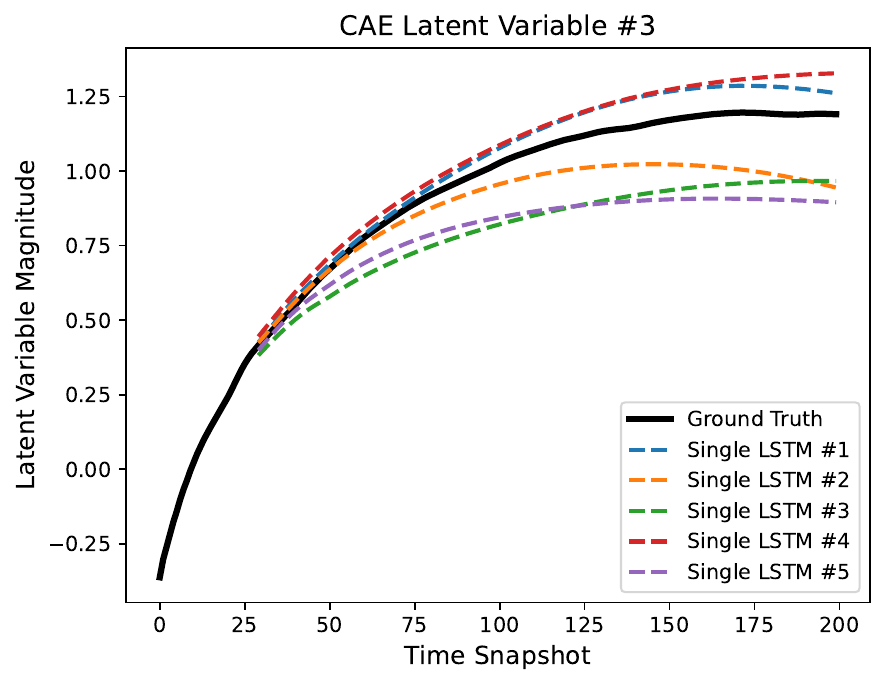}
  \includegraphics[width=0.42\textwidth]{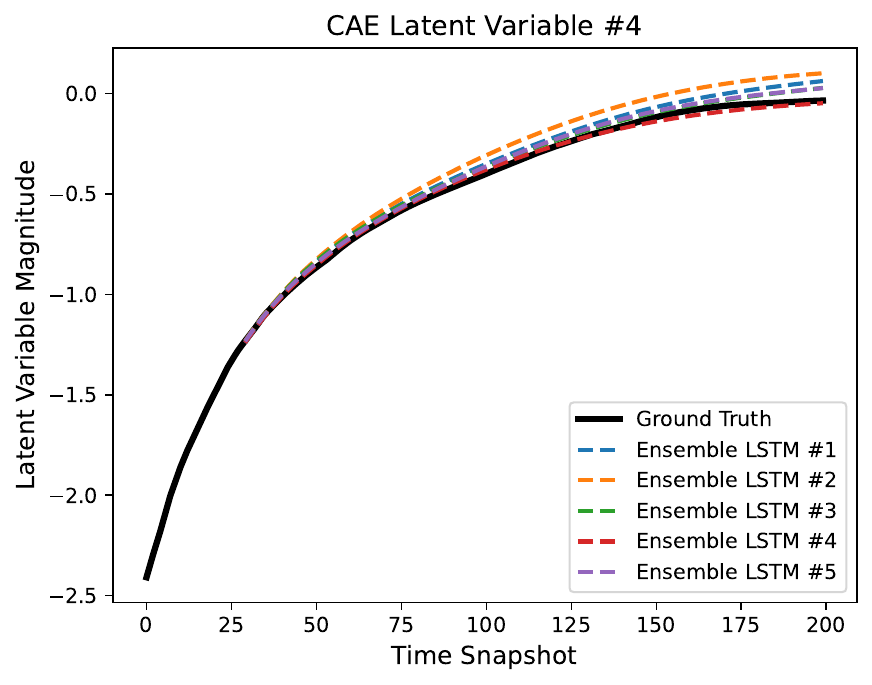}
  \includegraphics[width=0.42\textwidth]{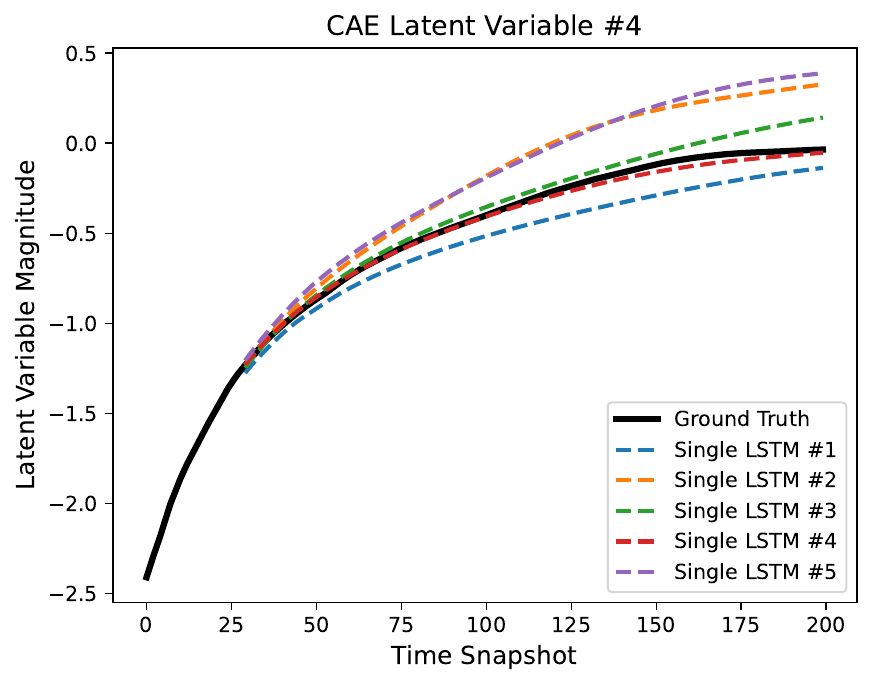}
  \caption{Comparison of latent variable trajectories for the lid-driven cavity case at $\bm{\mu^{*}} = [1.299, 1.689, -0.0367]$.}
  \label{fig:cavity_latent}
  \end{figure}

\begin{figure}[!h]
  \centering
  \includegraphics[width=0.25\textwidth]{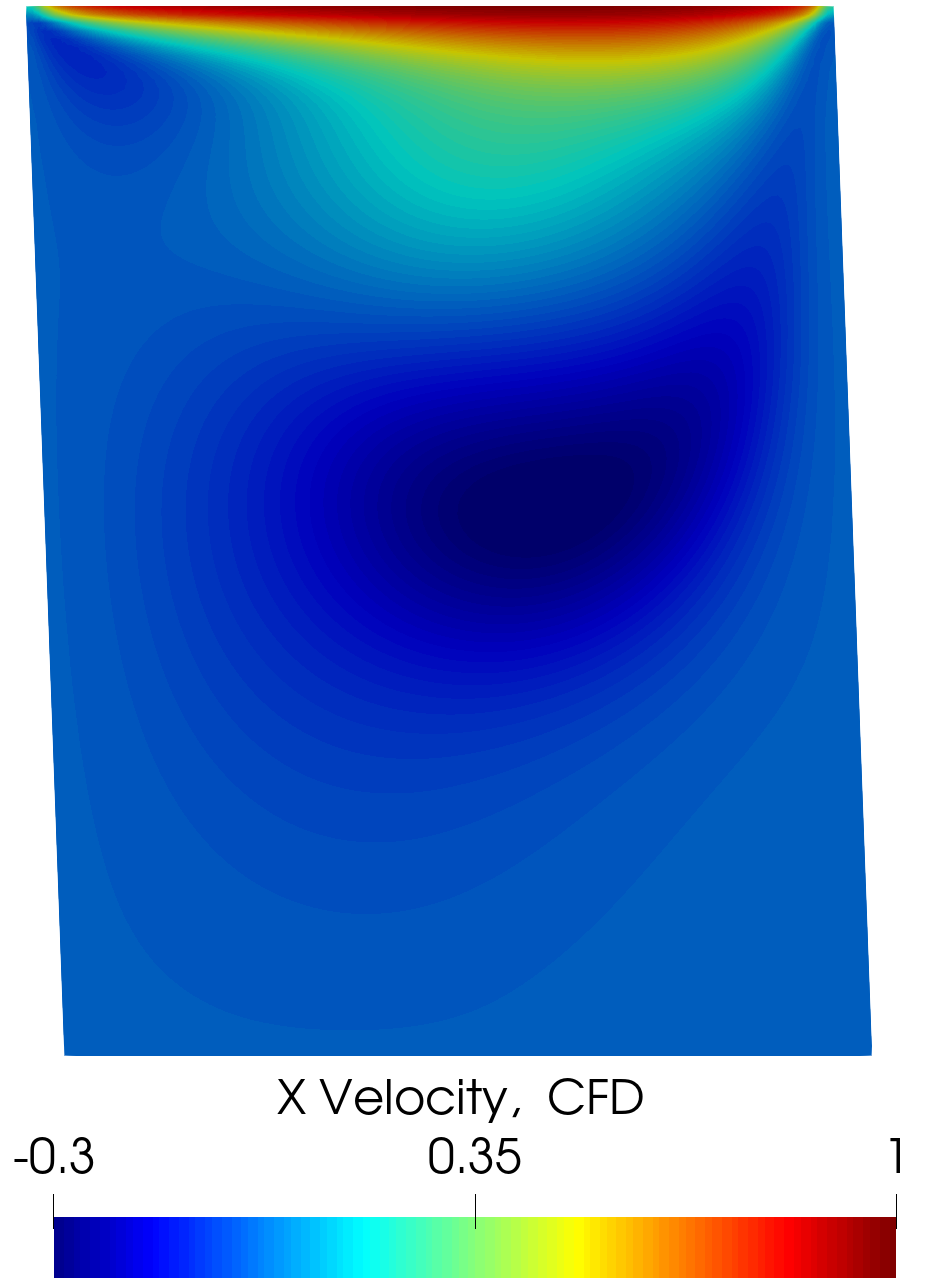}
  \includegraphics[width=0.25\textwidth]{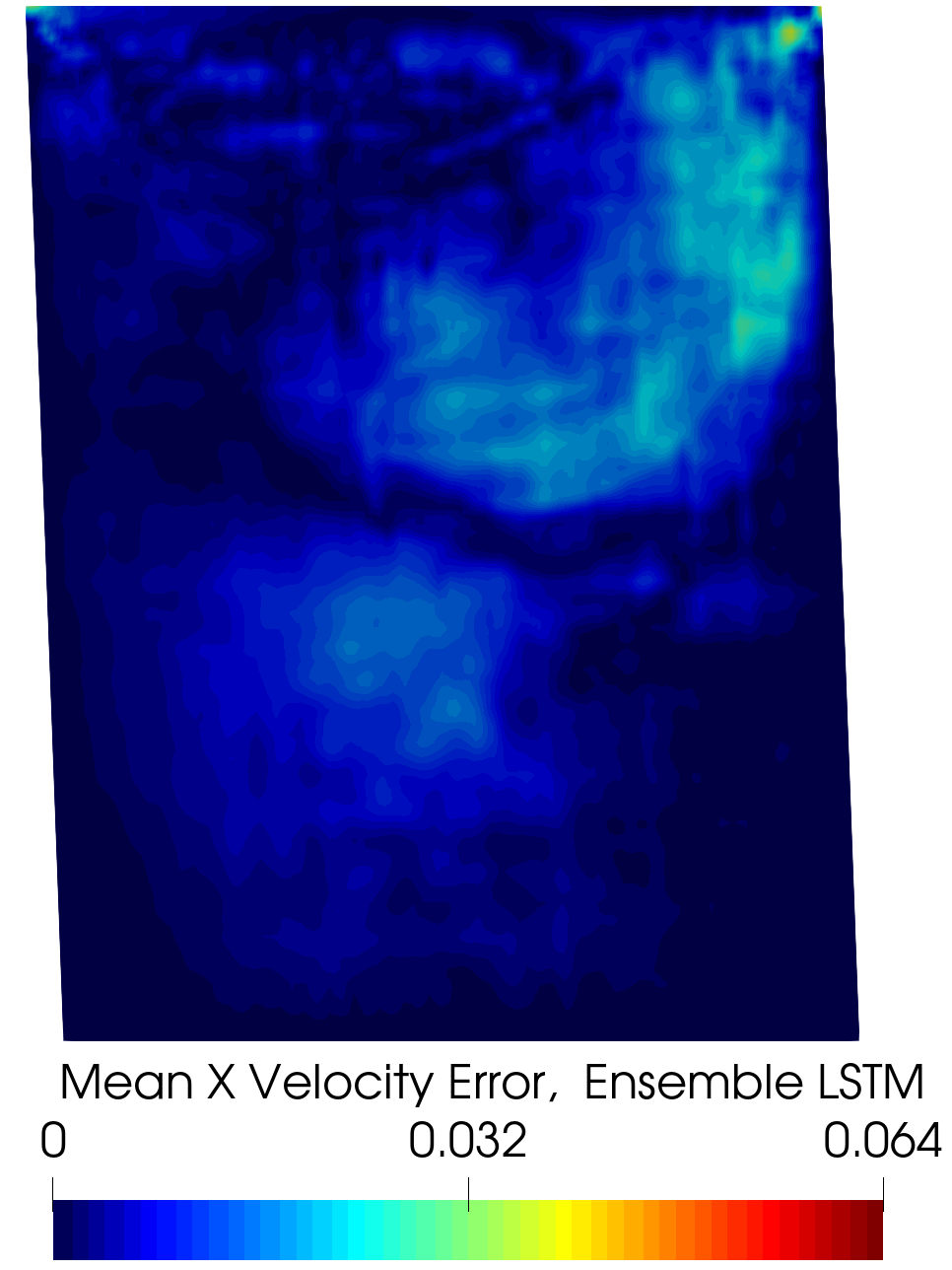}
  \includegraphics[width=0.25\textwidth]{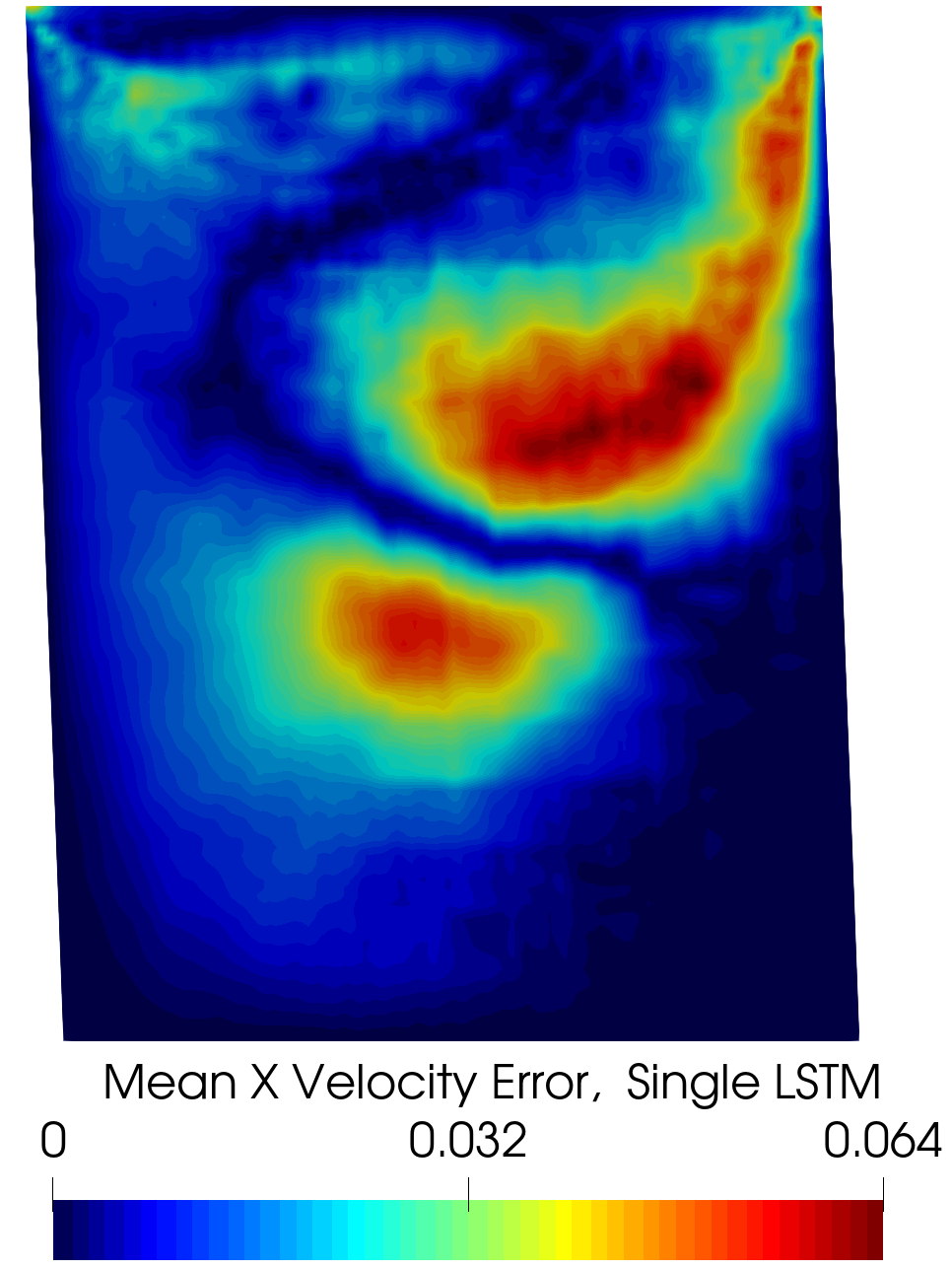}
  \includegraphics[width=0.25\textwidth]{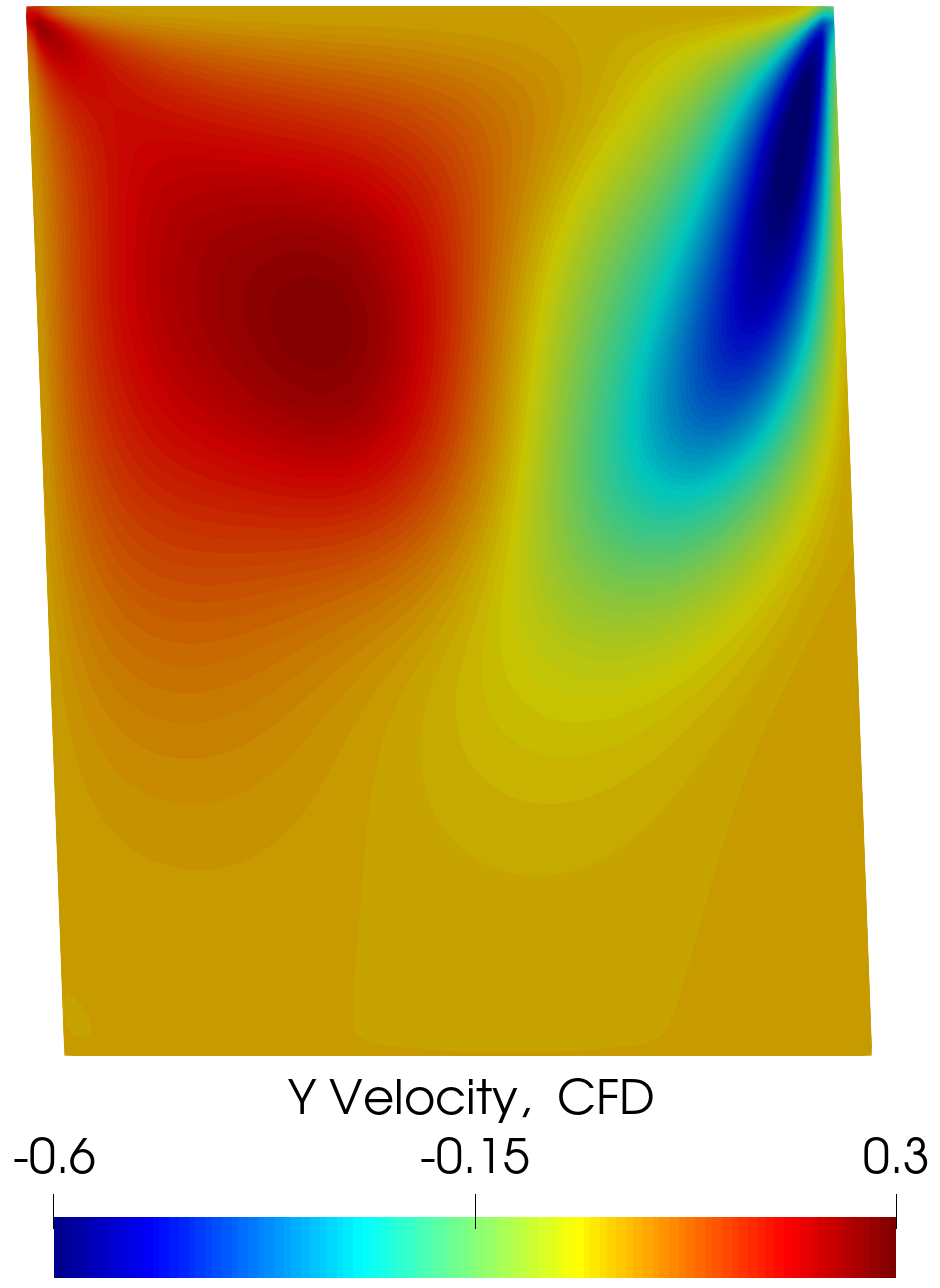}
  \includegraphics[width=0.25\textwidth]{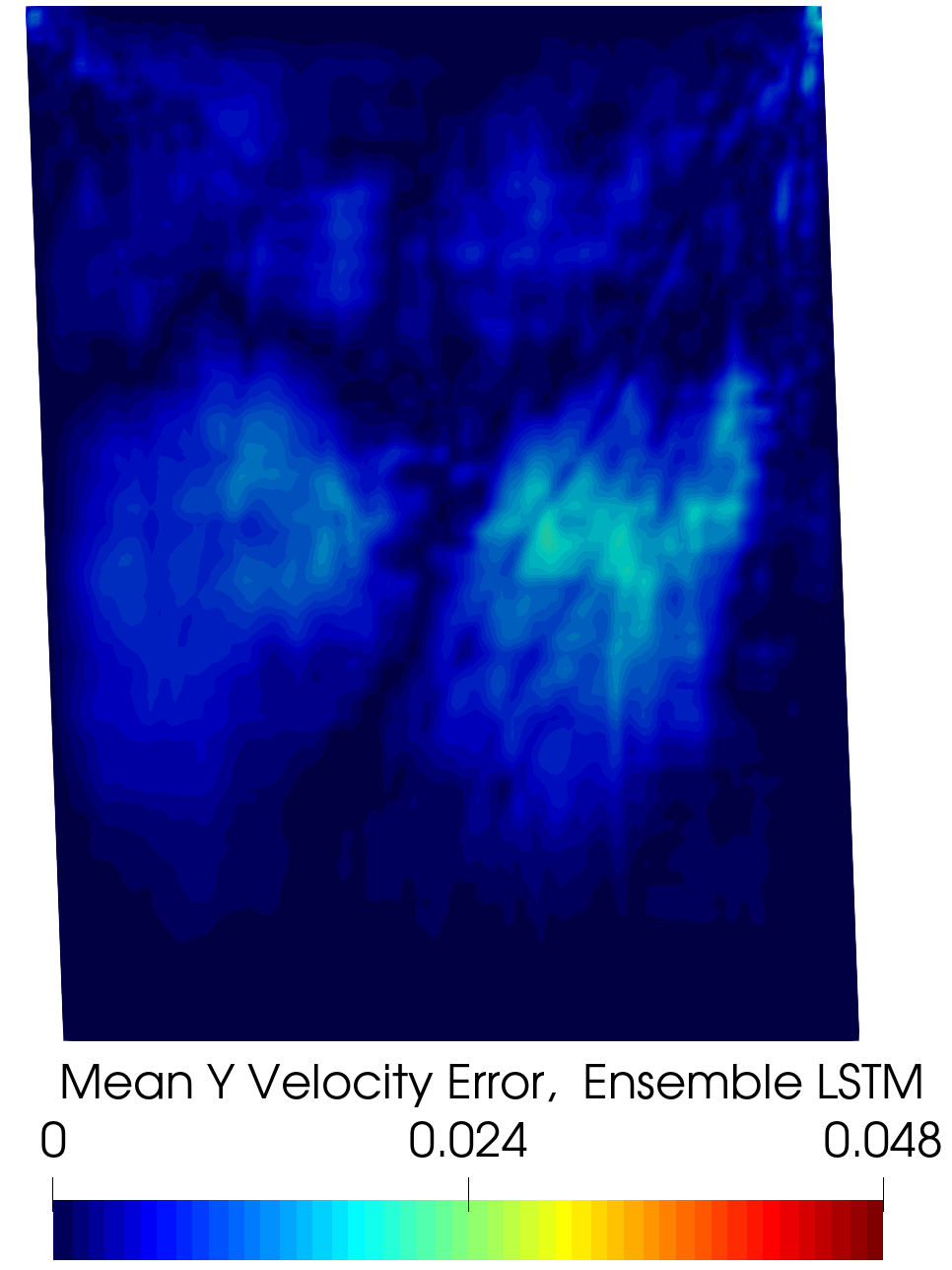}
  \includegraphics[width=0.25\textwidth]{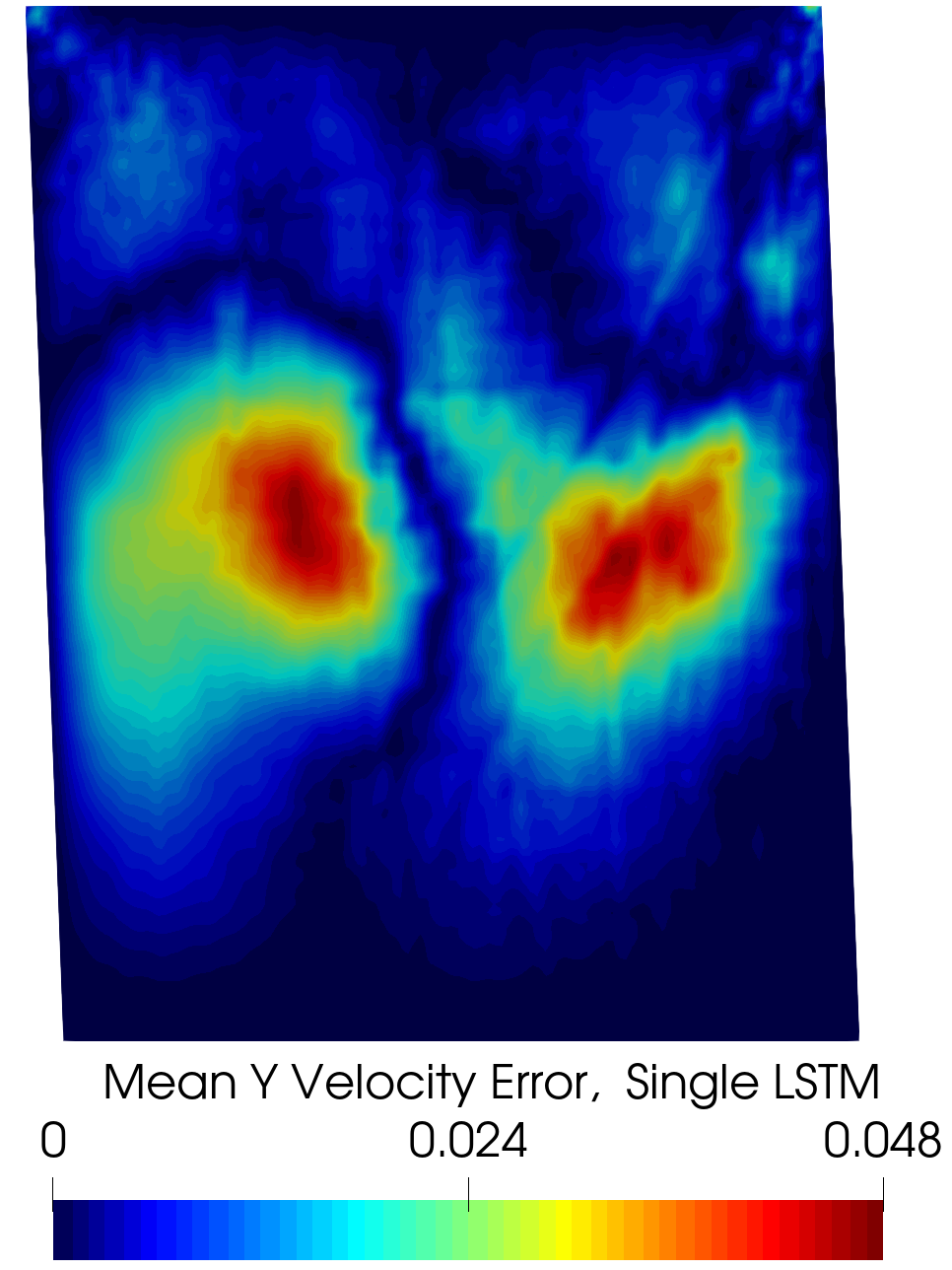}
  \caption{Time snapshot of $u$ (top) and $v$ (bottom) at T = 5 and errors averaged over the last 10\% of the simulation at $\bm{\mu^{*}} = [1.299, 1.689, -0.0367]$.}
  \label{fig:cavity_err_contours}
  \end{figure}

Figure~\ref{fig:cavity_latent} shows the trajectories of the latent variables computed using ensemble and single LSTMs at the test point $\bm{\mu^{*}} = [1.299, 1.689, -0.0367]$ using five different seeds which control the initial weights and biases of the LSTM. It is shown that for all of the latent variables, using LSTM ensembles leads to higher prediction accuracy and significantly lower variance. When using a single LSTM, the different predictions quickly diverge from each other and the ground truth. Due to their high capacity to learn, neural networks often exhibit high variance. Since neural networks contain a large number of parameters, the effects of error propagation between different trained models are heightened due to relatively large differences in autoregressive predictions being fed back into the model. In contrast, the ensemble model is much less sensitive to the initial seed, and the predicted trajectories do not differ much. The second latent variable exhibits diverging trajectories when using the ensemble model, but the effect is much less pronounced than when using a single LSTM. Figure~\ref{fig:cavity_err_contours} shows contours of $u$ and $v$ at the test point as well as absolute ROM errors averaged over the last 10\% of the simulation ($t \in$ [4.5, 5] seconds) for a single seed. The errors are considerably larger when using a single LSTM throughout the computational domain for both $u$ and $v$. Table~\ref{tab:cavity_errs} lists the seed-averaged relative errors $\bar{\epsilon}$ in $u$ and $v$ for the test points. The ensemble method offers better performance in predicting both fields at all of the test points, usually by wide margins. Table~\ref{tab:cavity_std} lists the standard deviation of these errors; at all of the test points, the standard deviation is also smaller. As there is a lower spread amongst the different seeds in the time-averaged error metrics, this shows that using the ensemble method leads to much greater stability and reliability in predictions.  Table~\ref{tab:cavity_costs} lists the computational costs associated with neural network training, CFD simulation, and ROM inference. The given CFD wall time is the portion of the simulation over the prediction time horizon; using the ROM for inference over this period of time offers a speed-up of approximately 9.4x over CFD. The ROM inference costs include autoregressive prediction using the ensemble LSTM and obtaining full-order approximations using the decoder. The CAE is much more expensive to train than a single LSTM as it has a larger number of trainable parameters. While a multi-GPU architecture was used to train the LSTMs in parallel, the cost to train a single LSTM is given.

\clearpage

\begin{table}[!h]
  \centering
  \begin{tabular}{@{}lllll@{}}
  \toprule
  Test Case Index & $\bar{\epsilon}$, $u$ (Ensemble) & $\bar{\epsilon}$, $u$ (Single) & $\bar{\epsilon}$, $v$ (Ensemble) & $\bar{\epsilon}$, $v$ (Single) \\ \midrule
  1               & \textbf{0.02421}                                               & 0.07119                                                      & \textbf{0.02514}                                               & 0.07204                                                      \\
  2               & \textbf{0.01958}                                               & 0.04199                                                      & \textbf{0.02174}                                               & 0.03870                                                      \\
  3               & \textbf{0.02622}                                               & 0.05282                                                      & \textbf{0.02868}                                               & 0.06296                                                      \\
  4               & \textbf{0.03640}                                               & 0.06575                                                      & \textbf{0.03924}                                               & 0.06742                                                      \\
  5               & \textbf{0.02442}                                               & 0.06589                                                      & \textbf{0.01996}                                               & 0.05131                                                      \\
  6               & \textbf{0.04177}                                               & 0.04939                                                      & \textbf{0.03964}                                               & 0.04733                                                      \\
  7               & \textbf{0.02445}                                               & 0.03771                                                      & \textbf{0.02789}                                               & 0.03915                                                      \\
  8               & \textbf{0.04559}                                               & 0.05304                                                      & \textbf{0.04251}                                               & 0.05194                                                      \\
  9               & \textbf{0.03313}                                               & 0.04678                                                      & \textbf{0.03891}                                               & 0.04900                                                      \\
  10              & \textbf{0.02877}                                               & 0.04291                                                      & \textbf{0.02717}                                               & 0.04079                                                      \\ \bottomrule
  \end{tabular}
  \caption{Average relative errors in $u$ and $v$ for the lid-driven cavity case. Boldface denotes the lowest values.}
  \label{tab:cavity_errs}
  \end{table}

\begin{table}[!h]
  \centering
  \begin{tabular}{@{}lllll@{}}
  \toprule
  Test Case Index & $\sigma_s$, $u$ (Ensemble) & $\sigma_s$, $u$ (Single) & $\sigma_s$, $v$ (Ensemble) & $\sigma_s$, $v$ (Single) \\ \midrule
  1               & \textbf{0.005225}   & 0.02819           & \textbf{0.004470}   & 0.03051           \\
  2               & \textbf{0.001637}   & 0.009202          & \textbf{0.001766}   & 0.01267           \\
  3               & \textbf{0.004603}   & 0.01885           & \textbf{0.004040}   & 0.02263           \\
  4               & \textbf{0.008210}   & 0.02007           & \textbf{0.006915}   & 0.01172           \\
  5               & \textbf{0.003980}   & 0.01121           & \textbf{0.004222}   & 0.007739          \\
  6               & \textbf{0.003244}   & 0.01555           & \textbf{0.002547}   & 0.01726           \\
  7               & \textbf{0.002077}   & 0.007614          & \textbf{0.002466}   & 0.006886          \\
  8               & \textbf{0.002393}   & 0.02031           & \textbf{0.002030}   & 0.02041           \\
  9               & \textbf{0.002515}   & 0.007371          & \textbf{0.002550}   & 0.01324           \\
  10              & \textbf{0.002467}   & 0.01627           & \textbf{0.001759}   & 0.01493           \\ \bottomrule
  \end{tabular}
  \caption{Standard deviation in relative errors of $u$ and $v$ for the lid-driven cavity case. Boldface denotes the lowest values.}
  \label{tab:cavity_std}
  \end{table}
  
\begin{table}[!h]
  \centering
  \begin{tabular}{@{}ll@{}}
  \toprule
  Task         & Wall Time (s) \\ \midrule
  CAE           & 749           \\ 
  Single LSTM   & 81            \\ 
  CFD (1 CPU)   & 136          \\ 
  ROM Inference & 14.5          \\ \bottomrule
  \end{tabular}
  \caption{Computational costs associated with the lid-driven cavity problem.}
  \label{tab:cavity_costs}
  \end{table}


\subsection{2D Cylinder}

The next test case involves two-dimensional incompressible, unsteady, laminar flow over a cylinder between two solid walls. Eventually, the lid-driven cavity flow from the previous test case reaches steady-state and does not exhibit long-term transient behavior that is commonly found in fluid dynamics problems. Laminar flow over a cylinder is a well-studied problem in fluid dynamics, with both experimental and computational results present in the literature~\cite{tritton1959experiments,saiki1996numerical}. Unsteady cylinder flow is characterized by the presence of vortices that separate from the surface and form in the wake. This distinctive pattern is known as the Von Kármán vortex street, where alternating vortices of a regular pattern are shed downstream of the cylinder. The unsteady Navier-Stokes equations found in equations~\ref{eqn_mass} and~\ref{eqn_momentum} are solved using XLB~\cite{ataei2024xlb}, a Lattice Boltzmann method~\cite{chen1998lattice} (LBM) library utilizing the JAX framework~\cite{jax2018github} available for Python, which allows for effective scaling across multiple CPUs, GPUs, and distributed multi-GPU systems. The cited works can be referred to for an overview of the Lattice Boltzmann method and its implementation in XLB. 

No-slip boundary conditions are applied to the cylinder's surface and top and bottom walls. A Poiseuille flow profile is used for the inlet velocity. Extrapolation outflow boundary conditions are used for the outlet to allow the fluid flow to exit the domain freely.  The computational domain measures 1536 $\times$ 512 voxels that are uniformly spaced. The cylinder is centered at $x_{c}, y_{c} = [160, 256]$ (zero-based indexing is used). Simulation results are down-sampled onto a grid that measures 384 $\times$ 128 before being used for the ROM, resulting in $N = 49152$, as the original domain's large size leads to a very large training cost as well as memory usage. Two design parameters are used, the diameter $d$ of the cylinder and the Reynolds number $Re$. The bounds of the design parameters are given as 

\begin{align*}
  & \mu_{1} = d \in [48, 68], \\
  & \mu_{2} = Re \in [120, 240].
  \end{align*}

The diameter $d$ is set to an integer quantity by rounding to the nearest whole number. The wake structure behind the cylinder undergoes instabilities~\cite{gerrard_1997,rajani2009numerical} at a critical reynolds number of approximately $Re_{c} = 180$, where the vortices transition to becoming turbulent. As a result, the prescribed range given for $Re$ in the parameter space includes a variety of physical regimes. Additionally, both $Re$ and $d$ control the size and periodicity of the shed vortices, making this a difficult prediction problem for ROMs. The freestream velocity in the $x$-direction is set to $u_\infty  = 0.001$ in non-dimensional units. The simulation is run for $T = 750000$ timesteps with data being saved every 1000 steps, leading to 750 snapshots for a single simulation. In LBM simulations, time and space are often non-dimensionalized by $\Delta t $ and $\Delta x$ to arrive at dimensionless values of $\Delta x^* = \Delta t^* = 1$. However, one needs to ensure that $\Delta t / \Delta x \ll 1$ to avoid compressibility errors. In these runs we picked $\Delta t / \Delta x = 0.001$. The initial condition for a simulation of a given diameter $d$ is the solution to steady flow at $Re = 20$. 

\begin{figure}[!h]
\centering
\includegraphics[width=0.5\textwidth]{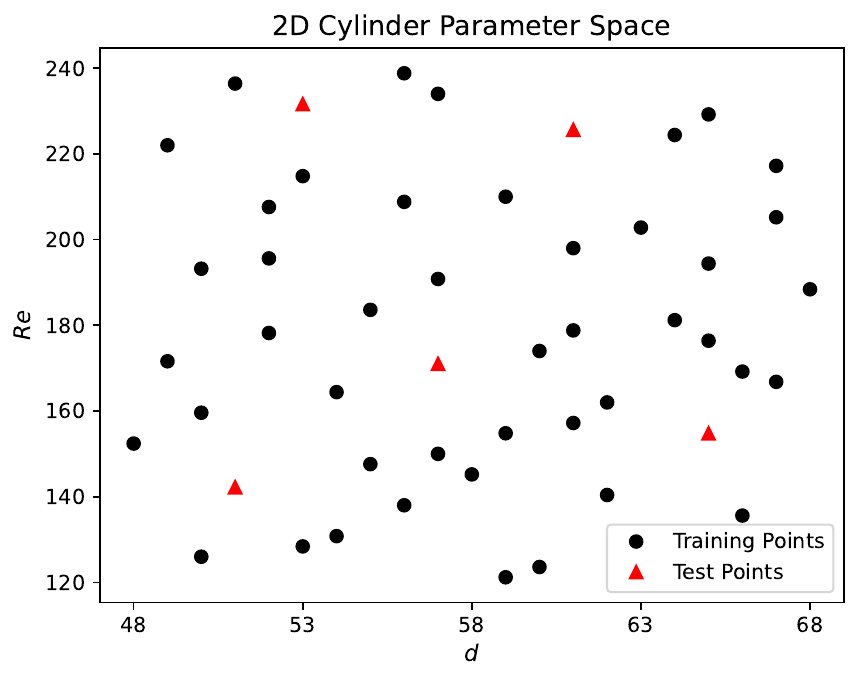}
\caption{Training and test design parameters for the 2D cylinder case.}
\label{fig:cyl_dv}
\end{figure}

50 sets of design parameters are generated and split into 45 training samples and 5 test points, shown in Figure~\ref{fig:cyl_dv}. Table~\ref{tab:cylinder_dv} also lists the test point design parameters. The ROM uses $m = 96$ bagged LSTMs and a window size of $w$ = 30, which were again chosen through a trial-and-error process to maximize accuracy while keeping the number of weak learners to a minimum. The CAE latent dimension is set to $k = 10$. Similar to the lid-driven cavity case, values below $k = 10$ resulted in higher reconstruction errors and increasing the latent dimension further offers no improvement. A bidirectional LSTM architecture~\cite{schuster1997bidirectional} is used. Bidirectional recurrent neural networks process sequential data in both forward and backward directions, allowing the model to learn both past and future context. For problems that are characterized by regularly repeating patterns such as a vortex street, bidirectional LSTMs can better learn the cyclic nature of the pattern by leveraging context in both directions. The network consists of three hidden layers with 36 neurons each and a dropout rate of 0.15. The output layer again contains a sigmoid activation function, and the Adam optimizer with the same initial learning rate $\eta = 5 \times 10^{-4}$ and weight decay of $\lambda = 1 \times 10^{-6}$ is used for both the LSTM and CAE. Again, the CAE is trained for 200 epochs while an individual LSTM is trained for 250 epochs. The CAE architecture is given in Appendix~\ref{appendix:arch}. At the test points, the FOM is run for $T_{i} = 300000$ seconds (300 snapshots), or 40\% of the total simulation time. This value is required to be high as the flow exhibits highly oscillatory behavior initially, leading to very noisy latent variables that cannot be used for model training. As a result, latent variable sequences are generated starting at the 200th snapshot, and  $\bm{A}_{\text{train}}$ does not contain time-series data of latent variables before this point.

\begin{figure}[!h]
  \centering
  \includegraphics[width=0.42\textwidth]{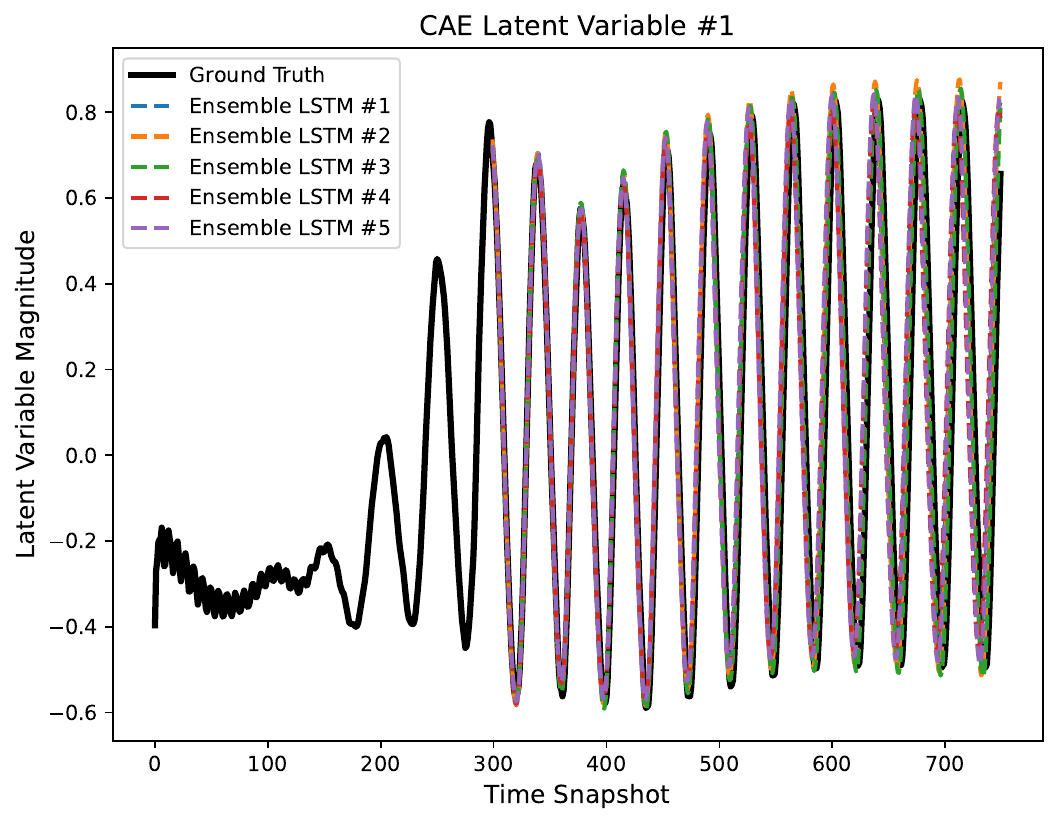}
  \includegraphics[width=0.42\textwidth]{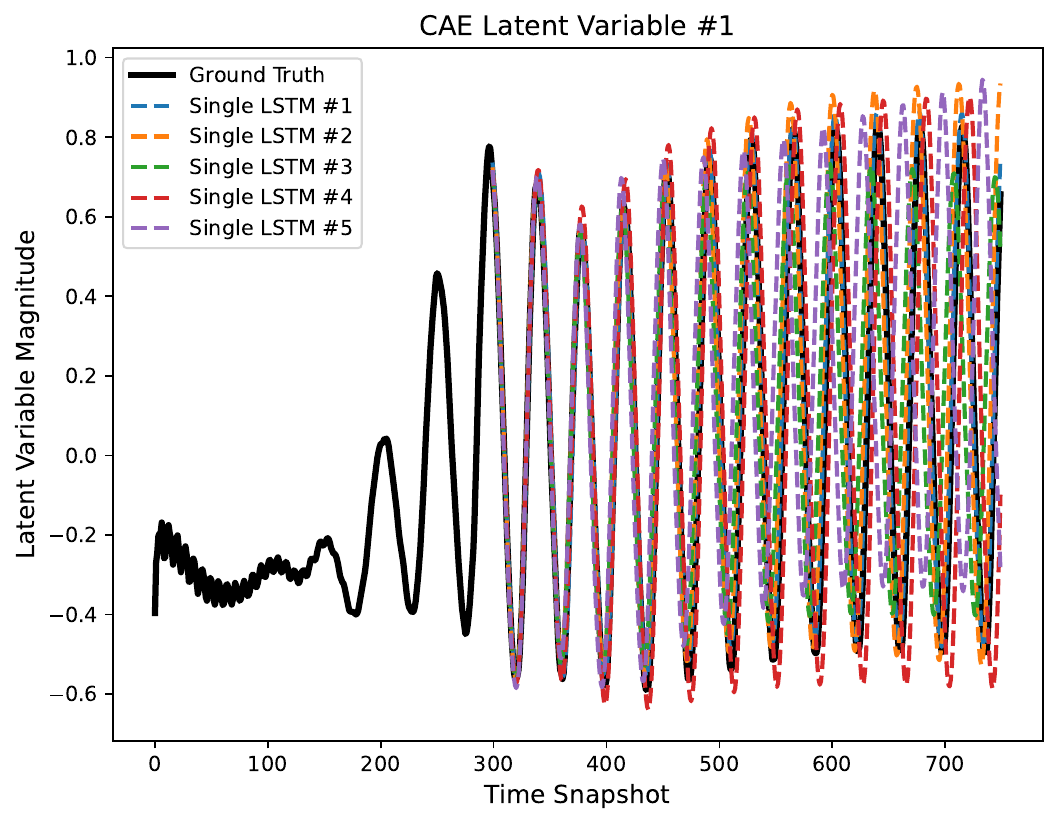}
  \includegraphics[width=0.42\textwidth]{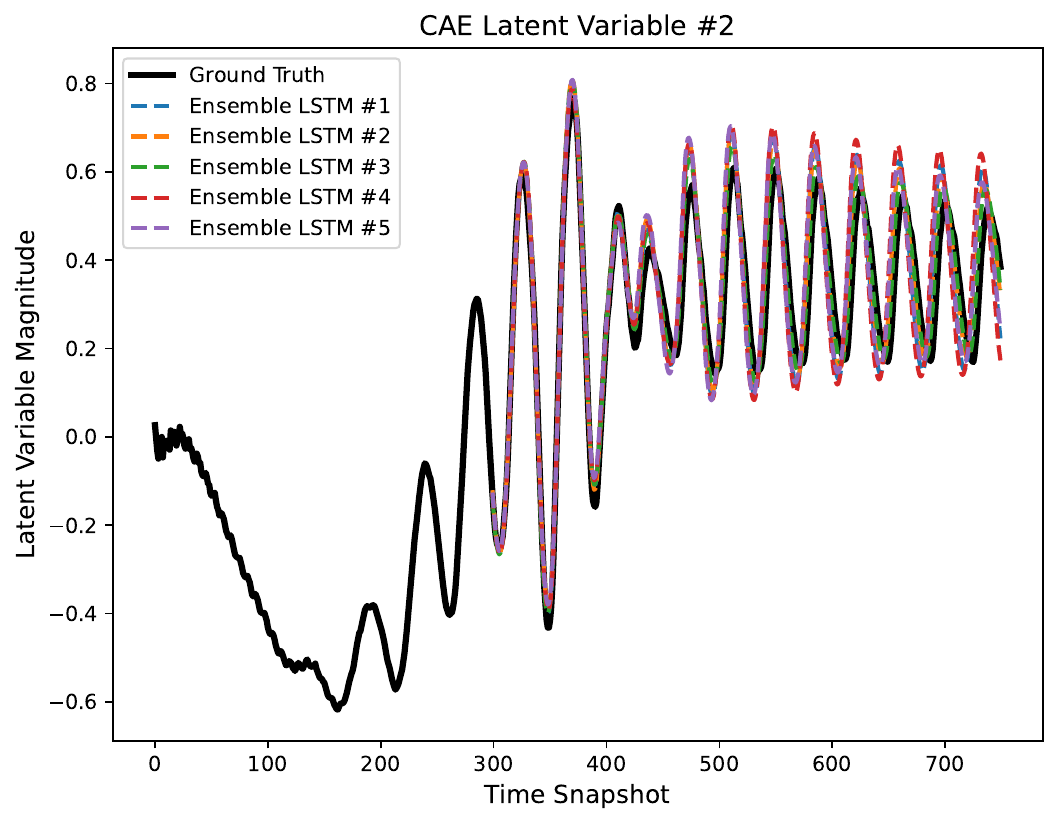}
  \includegraphics[width=0.42\textwidth]{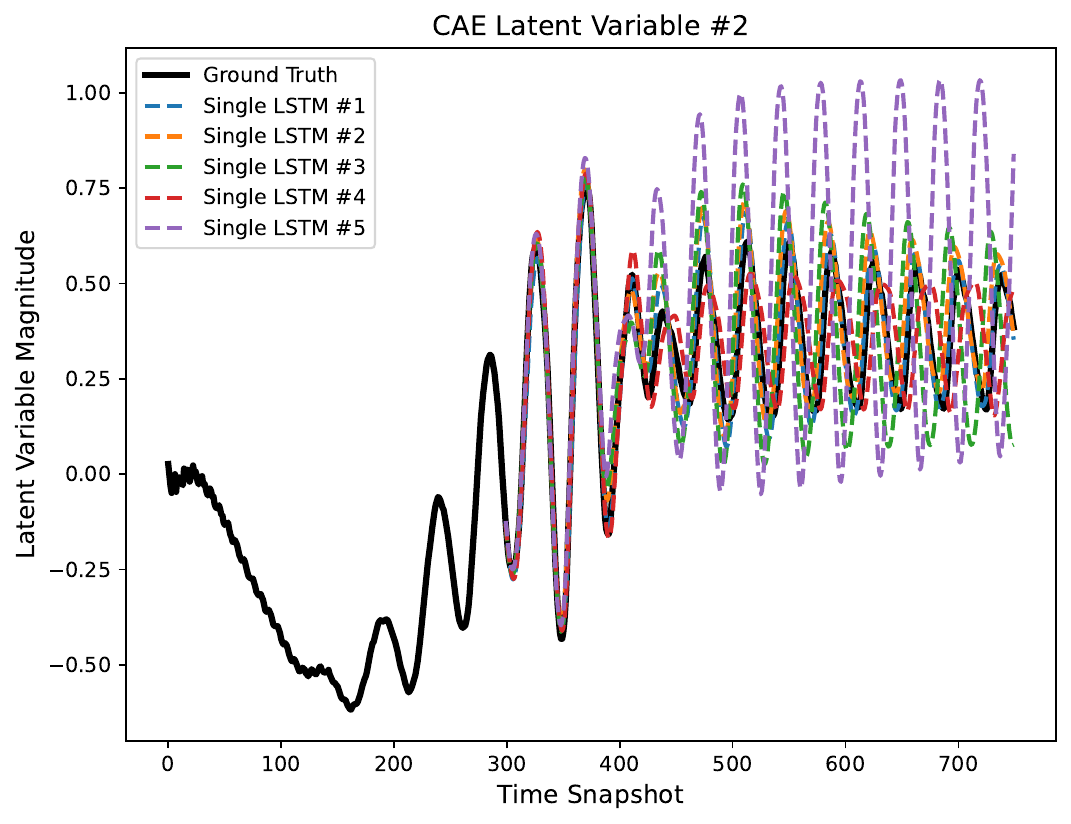}
  \includegraphics[width=0.42\textwidth]{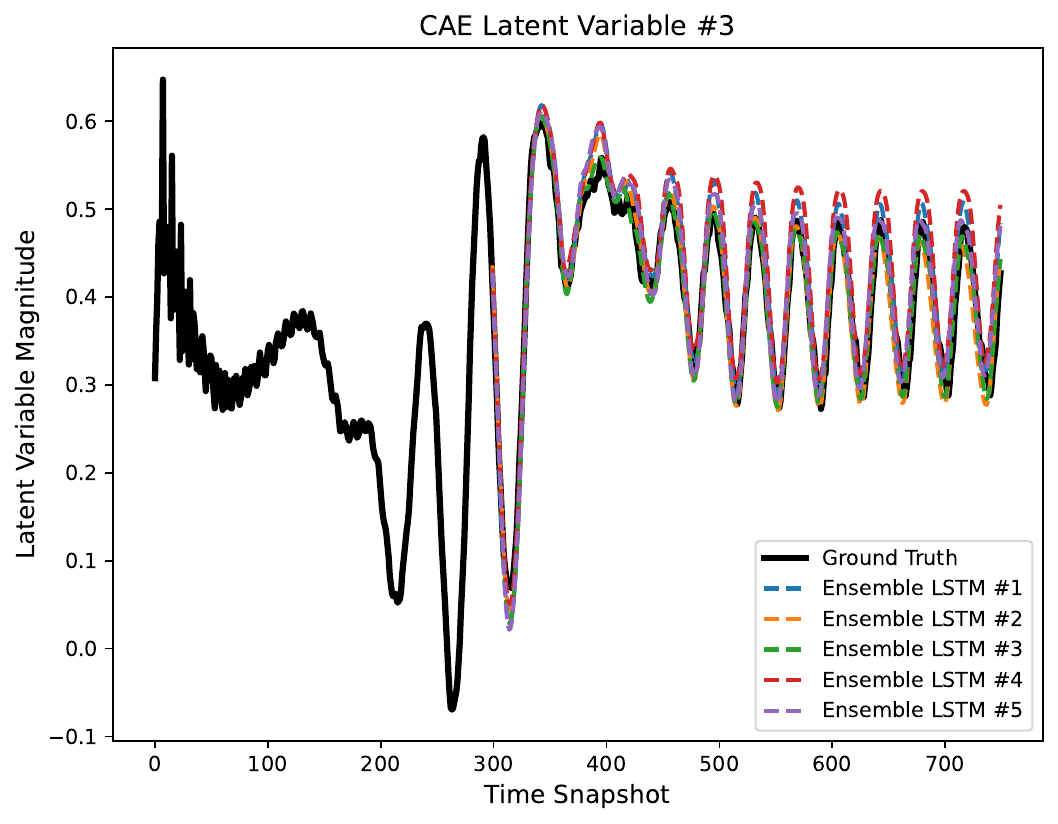}
  \includegraphics[width=0.42\textwidth]{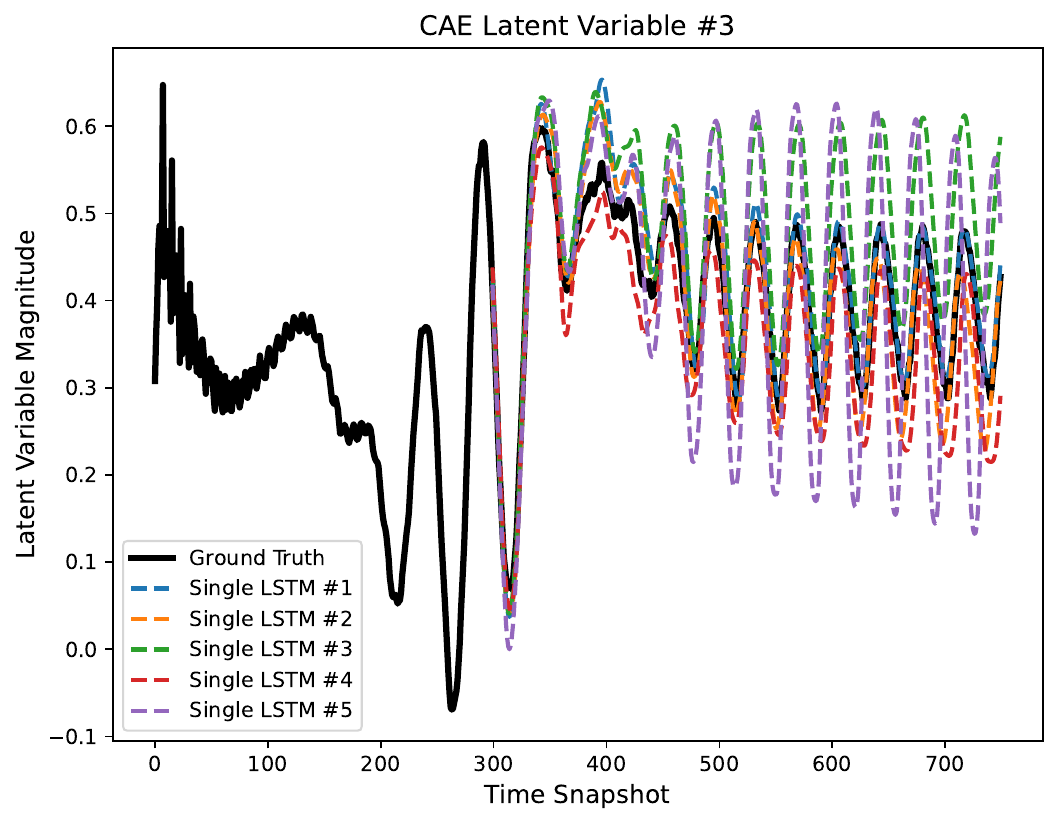}
  \includegraphics[width=0.42\textwidth]{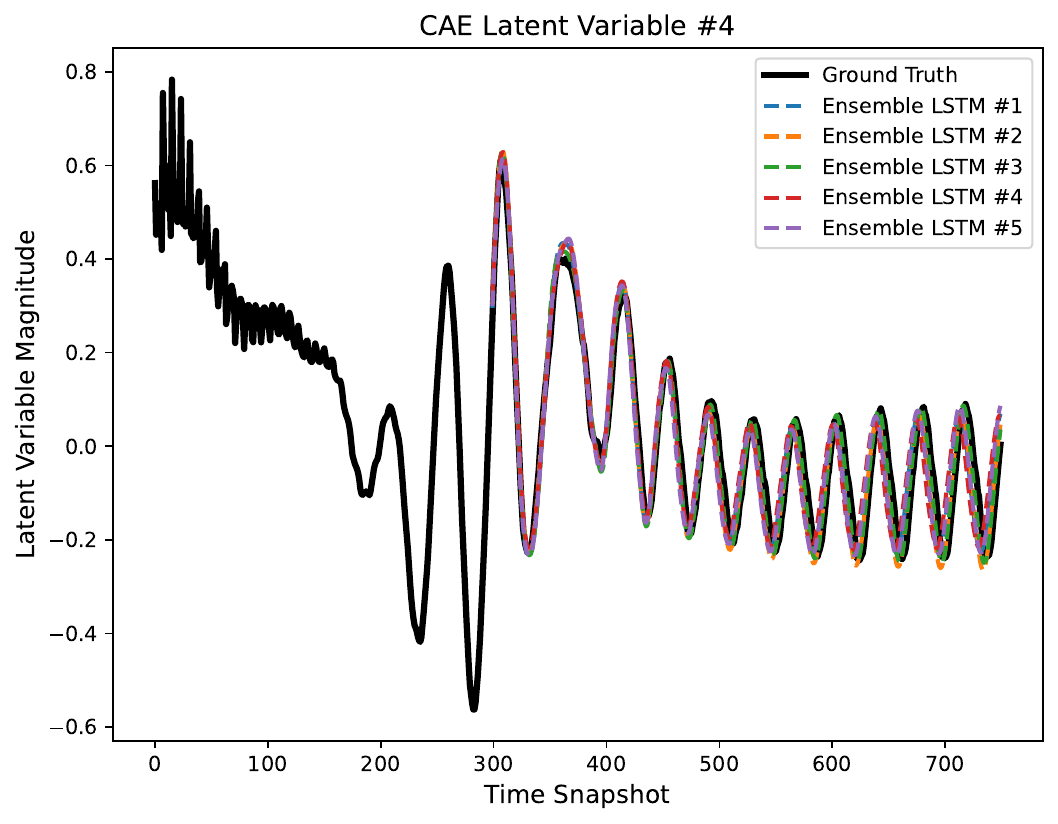}
  \includegraphics[width=0.42\textwidth]{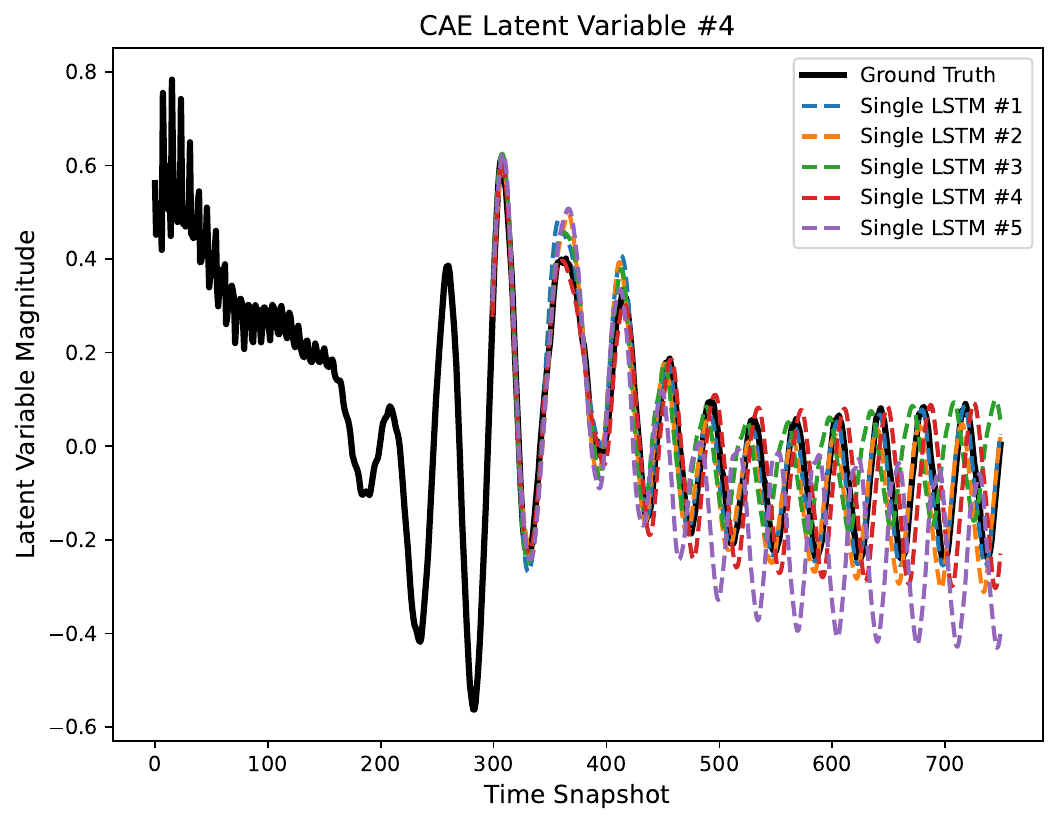}
  \caption{Comparison of latent variable trajectories for the 2D cylinder case at $\bm{\mu^{*}} = [51, 142.2]$.}
  \label{fig:cylinder_latent}
  \end{figure}

\clearpage

\begin{table}[!h]
  \centering
  \begin{tabular}{@{}lll@{}}
  \toprule
  Test Case Index & $d$   & $Re$  \\ \midrule
  1               & 53 & 231.6 \\ 
  2               & 51 & 142.2 \\ 
  3               & 65 & 154.8 \\ 
  4               & 57 & 171.0 \\ 
  5               & 61 & 225.6 \\ \bottomrule
  \end{tabular}
  \caption{Test case design parameters for the 2D cylinder case.}
  \label{tab:cylinder_dv}
  \end{table}

Figure~\ref{fig:cylinder_latent} shows the latent variable trajectories for the first four latent variables at the test point $\bm{\mu^{*}} = [51, 142.2]$. For each latent variable, the predictions given by single LSTMs are initially similar, but eventually diverge in terms of both the amplitude and frequency of the latent variables, leading to large inaccuracies. Using LSTM ensembles greatly reduces this effect, and the resultant latent variable trajectories follow the ground truth closely and do not differ greatly in amplitude nor frequency. The regular pattern of the Kármán vortex street is well-predicted given a small amount of initial latent variable history.

\begin{figure}[!h]
\centering
\includegraphics[width=0.49\textwidth]{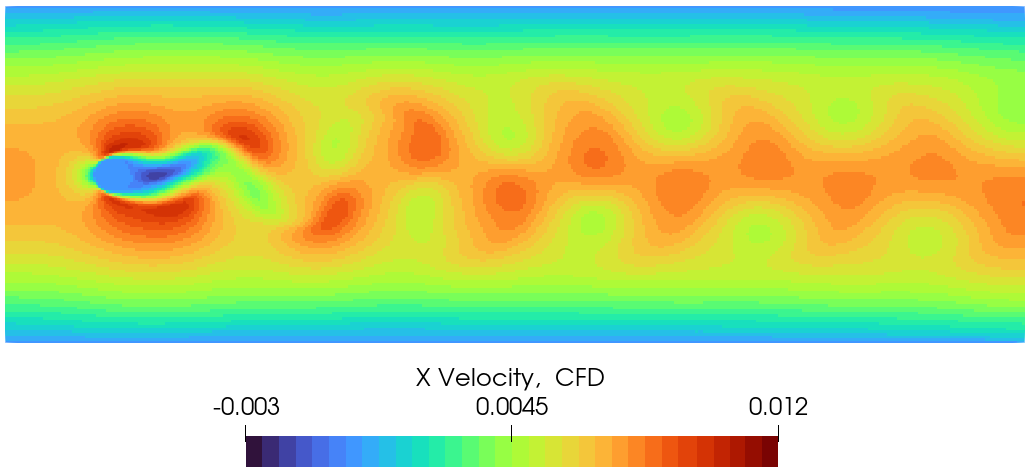}

\includegraphics[width=0.49\textwidth]{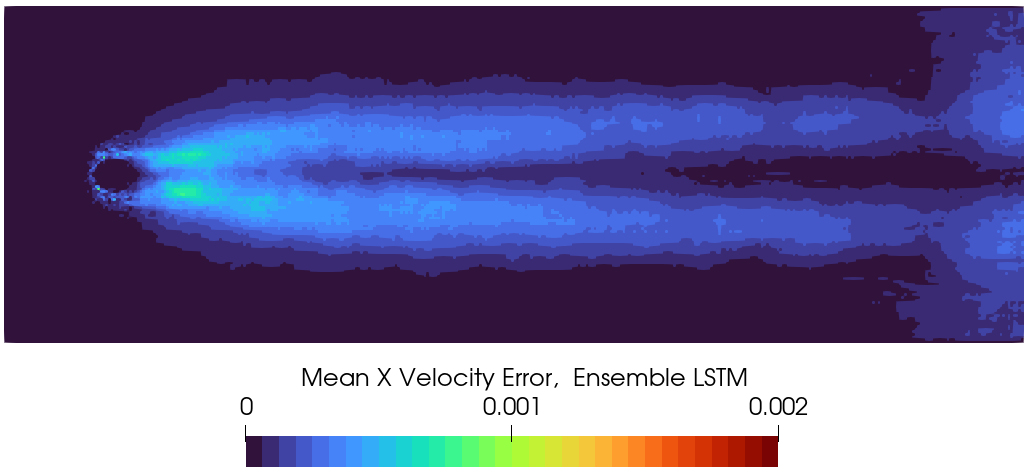}
\includegraphics[width=0.49\textwidth]{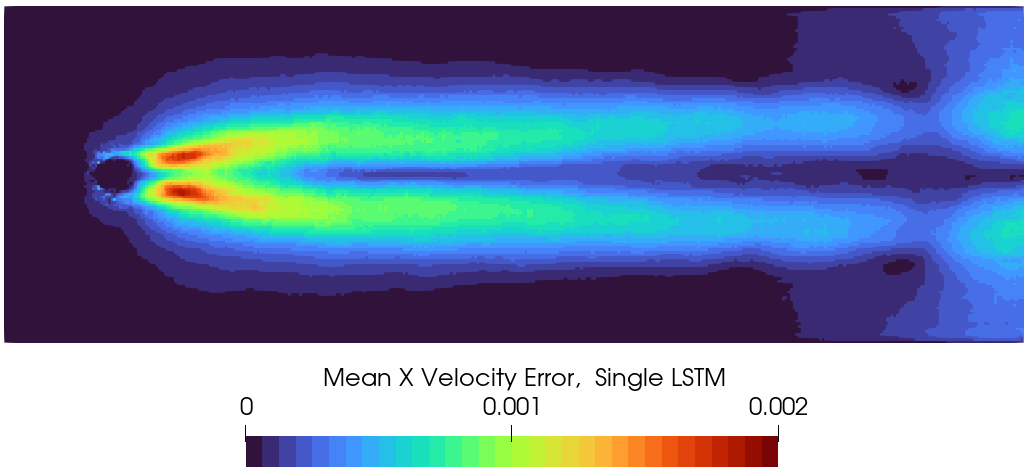}
\includegraphics[width=0.49\textwidth]{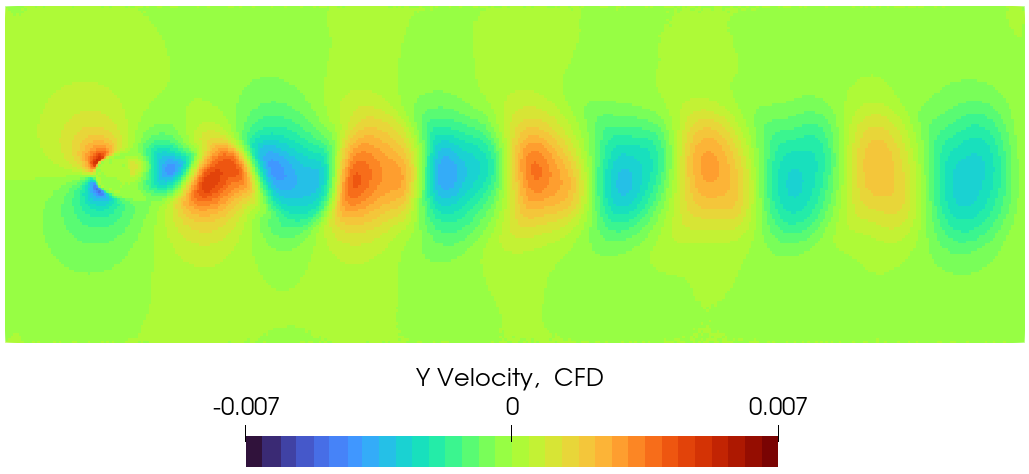}

\includegraphics[width=0.49\textwidth]{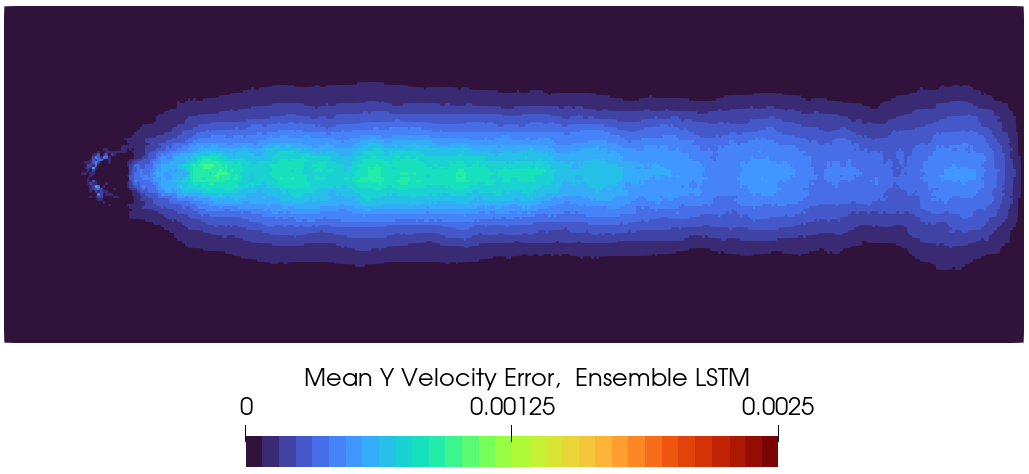}
\includegraphics[width=0.49\textwidth]{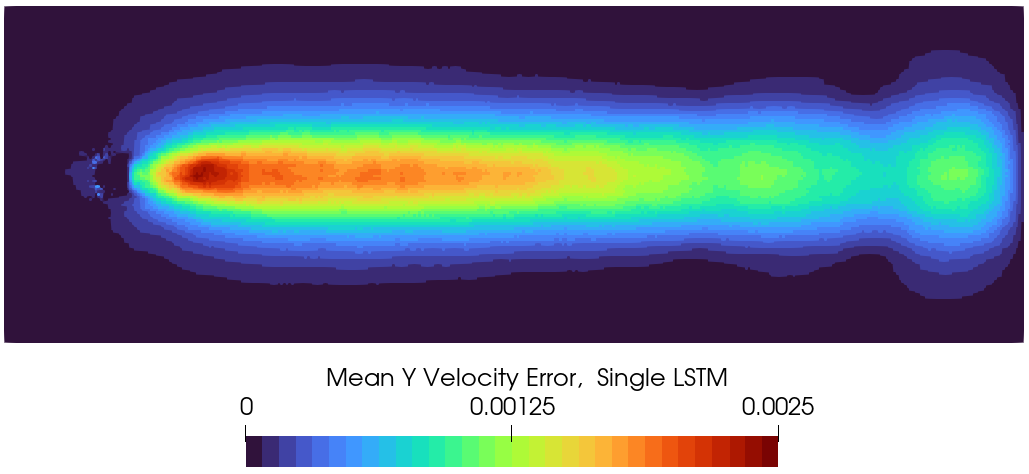}
\caption{Time snapshots of $u$ and $v$ at T = 750000 and errors averaged over the last 10\% of the simulation at $\bm{\mu^{*}} = [51, 142.2]$.}
\label{fig:cylinder_err_contours}
\end{figure}

Figure ~\ref{fig:cylinder_err_contours} shows snapshots of $u$ and $v$ at $T = 750000$ as well as ROM errors averaged over the last 10\% of the simulation for a single seed. The errors are significantly lower throughout the computational domain when using LSTM ensembles, which are greatest in the wake of the cylinder and immediately downstream of it. Table~\ref{tab:cylinder_errs} lists the seed-averaged relative errors in $u$ and $v$ at the test points. The ensemble ROM again offers better performance in predicting both fields at all of the test points. Table~\ref{tab:cylinder_std} lists the standard deviation of these errors, which are again lower at all of the test points. At the fourth test case index ($\bm{\mu^{*}} = [57, 171.0]$), the errors and standard deviations are similar. This test point lies in the middle of the parameter space and is well-surrounded by training samples; as a result, the advantage gained using an ensemble method may be marginal. Table~\ref{tab:cylinder_costs} lists the computational costs associated with the problem. Using the ROM for inference offers a speed-up of approximately 27x over CFD (again, the given CFD and ROM inference costs are for the portion of the simulation over the prediction time horizon). Due to the increased domain size and amount of training data, the CAE and LSTM are more expensive to train when compared to the lid-driven cavity case. 

    \begin{table}[!h]
      \centering
      \begin{tabular}{@{}lllll@{}}
      \toprule
      Test Case Index & $\bar{\epsilon}, u$ (Ensemble) & $\bar{\epsilon}, u$ (Single) & $\bar{\epsilon}, v$ (Ensemble) & $\bar{\epsilon}, v$ (Single) \\ \midrule
      1                        & \textbf{0.04324}               & 0.07975                      & \textbf{0.3379}               & 0.6162                       \\
      2                        & \textbf{0.03463}               & 0.05838                      & \textbf{0.2971}               & 0.4871                       \\
      3                        & \textbf{0.02380}               & 0.05195                      & \textbf{0.1399}               & 0.3315                       \\
      4                        & \textbf{0.05446}               & 0.06636                      & \textbf{0.3995}               & 0.4900                       \\
      5                        & \textbf{0.03655}               & 0.06029                      & \textbf{0.2269}               & 0.4059                       \\ \bottomrule
      \end{tabular}
      \caption{Average relative errors in $u$ and $v$ for the 2D cylinder case. Boldface denotes the lowest values.}
      \label{tab:cylinder_errs}
      \end{table}
    
      \begin{table}[!h]
        \centering
        \begin{tabular}{@{}lllll@{}}
        \toprule
        Test Case Index & $\sigma_s$, $u$ (Ensemble) & $\sigma_s$, $u$ (Single) & $\sigma_s$, $v$ (Ensemble) & $\sigma_s$, $v$ (Single) \\ \midrule
        1               & \textbf{0.01321}                                               & 0.04184                                                      & \textbf{0.1138}                                                & 0.3372                                                       \\
        2               & \textbf{0.007607}                                              & 0.03304                                                      & \textbf{0.07005}                                               & 0.2720                                                       \\
        3               & \textbf{0.004610}                                              & 0.01587                                                      & \textbf{0.03493}                                               & 0.1136                                                       \\
        4               & \textbf{0.01521}                                               & 0.02026                                                      & \textbf{0.1186}                                                & 0.1536                                                       \\
        5               & \textbf{0.007528}                                              & 0.03398                                                      & \textbf{0.05904}                                               & 0.2684                                                       \\ \bottomrule
        \end{tabular}
      \caption{Standard deviation in relative errors of $u$ and $v$ for the 2D cylinder case. Boldface denotes the lowest values.}
      \label{tab:cylinder_std}
      \end{table}

    \begin{table}[!h]
      \centering
      \begin{tabular}{@{}ll@{}}
      \toprule
      Task          & Wall Time (s) \\ \midrule
      CAE           & 2933          \\ 
      Single LSTM   & 140           \\ 
      CFD (2 GPUs)  & 1305          \\ 
      ROM Inference & 48.4          \\ \bottomrule
      \end{tabular}
      \caption{Computational costs associated with the 2D cylinder case.}
      \label{tab:cylinder_costs}
      \end{table}

\clearpage

\section{Conclusion}

This work presents a fully data-driven ROM framework for time-dependent physics simulations using convolutional autoencoders and LSTM ensembles. The combined framework, referred to as the CAE-eLSTM ROM, is used to mitigate the issue of error propagation when performing autoregressive time-series prediction at unseen data sets, a common problem in data-driven CFD applications. The ROM obtains a low-dimensional solution manifold using convolutional autoencoders, a type of artificial neural network useful for reconstructing spatially distributed data. Using the encoder section of the autoencoder, the low-dimensional latent variables of the training data are used to generate multivariate time-series sequences that are used to train LSTMs, a type of recurrent neural network useful for modeling sequential data. Bagging, a popular ensemble learning technique, is used to train multiple base LSTMs on subsets of the training sequences sampled randomly with replacement. Ensemble learning is chosen as it is a useful tool for improving the stability and accuracy of machine learning methods. At unseen test points, the full-order model is run to obtain an initial sequence of latent variables through the encoder and autoregressive time-series prediction is used to predict the latent variables over a prescribed time horizon. The predicted latent variables are then passed through the decoder to obtain spatial reconstructions of the FOM at different points in time. 

When applied to two incompressible, unsteady, laminar fluid dynamics problems, the proposed ROM exhibits high levels of accuracy when compared to a ROM using a single LSTM model. The use of LSTM ensembles significantly diminishes the error propagation issue, with latent variable trajectories from different weight initializations showing low levels of variance. Error metrics show that using the ensemble ROM leads to better predictive performance at all test points. Furthermore, the fully data-driven nature of the ROM allows it to be applied to two different CFD solvers. Although the cost associated with training ensemble neural networks is high, bagging allows for models to be trained in parallel, and multi-GPU architectures can be used to significantly reduce the cost. The addition of an ensemble model for time-series predictions, while simple, greatly improves the performance of the ROM and does not require additional training data. A limitation of the presented framework is its significantly increased offline training cost. While this can be alleviated to a great degree by training weak learners in parallel, multi-GPU platforms are expensive and may not be readily available for practitioners. Additionally, the use of convolutional autoencoders limits the application of the ROM to structured grids unless a mesh interpolation method is used. Future work will focus on further developing the method to apply it to three-dimensional geometries, problems exhibiting turbulent flow, and more complex geometries involving unstructured meshes. Problems consisting of time-series datasets that exhibit non-smooth variations over long periods of time will be of particular interest, as they are more representative of real-world applications and can reap greater benefits from ensemble learning. Computationally efficient optimal hyperparameter selection methods will also be explored to lower the offline costs of training the ROM.

\section*{Author Declarations}
\subsection*{Conflict of Interest}
The authors have no conflicts to disclose.

\subsection*{Author Contributions}
\textbf{Rakesh Halder: } Conceptualization (equal), Formal Analysis (lead), Methodology (lead), Software (equal), Writing/Original Draft Preparation (lead) \\ 
\textbf{Mohammadmehdi Ataei: } Conceptualization (equal), Software (equal), Supervision (equal), Writing/Review \& Editing (equal) \\ 
\textbf{Hesam Salehipour: } Conceptualization (equal), Software (equal), Supervision (equal), Writing/Review \& Editing (equal) \\ 
\textbf{Krzysztof Fidkowski: } Writing/Review \& Editing (equal) \\ 
\textbf{Kevin Maki: } Writing/Review \& Editing (equal)

\section*{Data Availability Statement}
The data that support the findings of this study are available from the corresponding author upon reasonable request.

\clearpage

\begin{appendices}
\section{Reconstruction Errors}
\label{appendix:pod}
\begin{figure}[!h]
  \centering

  \includegraphics[width=0.33\textwidth]{figures/cylinder/u_cfd.png}

  \includegraphics[width=0.33\textwidth]{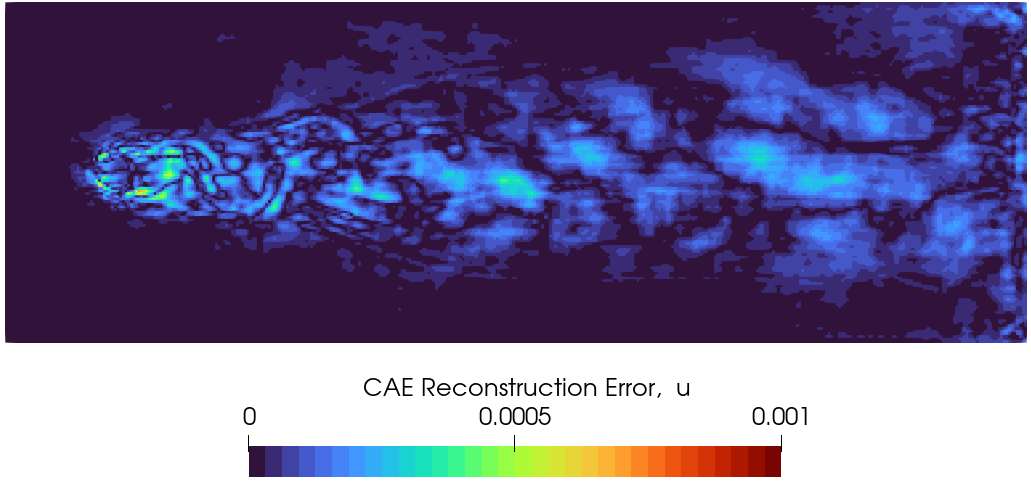}
  \includegraphics[width=0.33\textwidth]{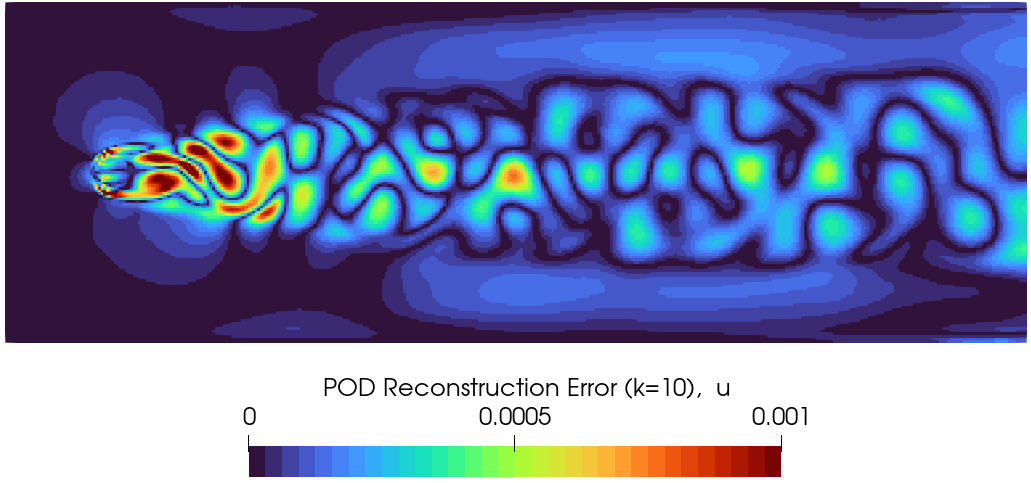}
  \includegraphics[width=0.33\textwidth]{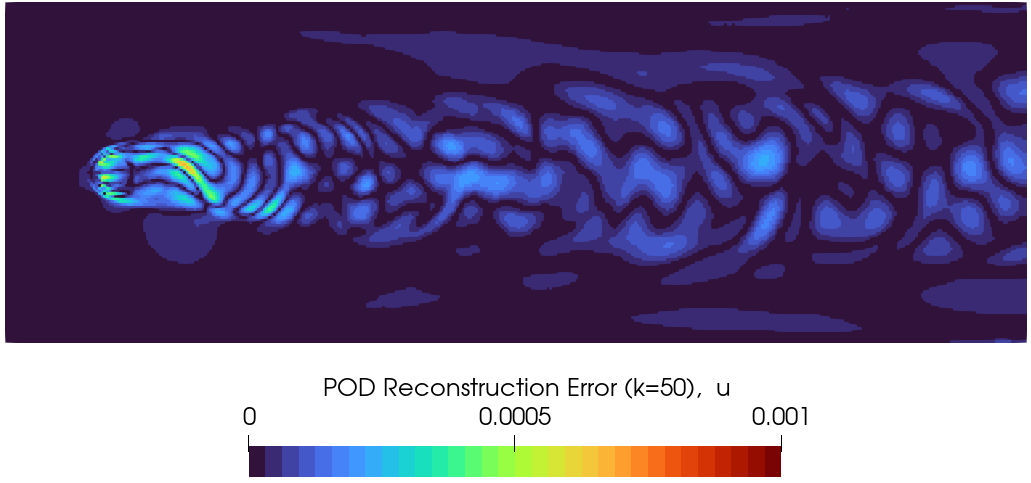}
  
  \includegraphics[width=0.33\textwidth]{figures/cylinder/v_cfd.png}
  
  \includegraphics[width=0.33\textwidth]{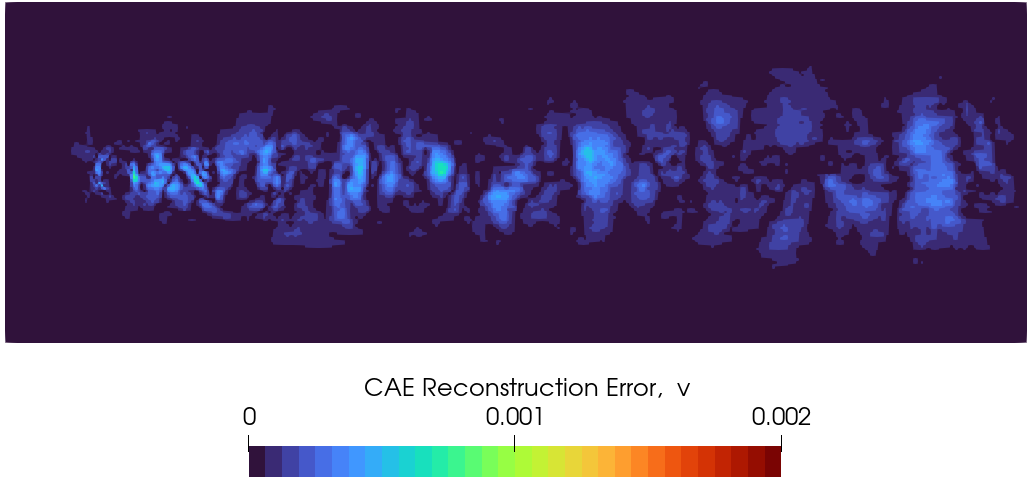}
  \includegraphics[width=0.33\textwidth]{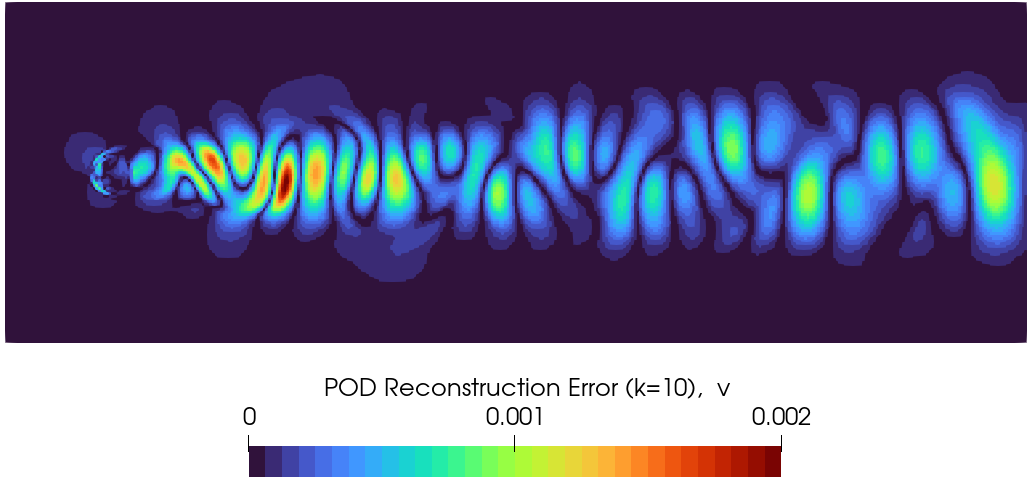}
  \includegraphics[width=0.33\textwidth]{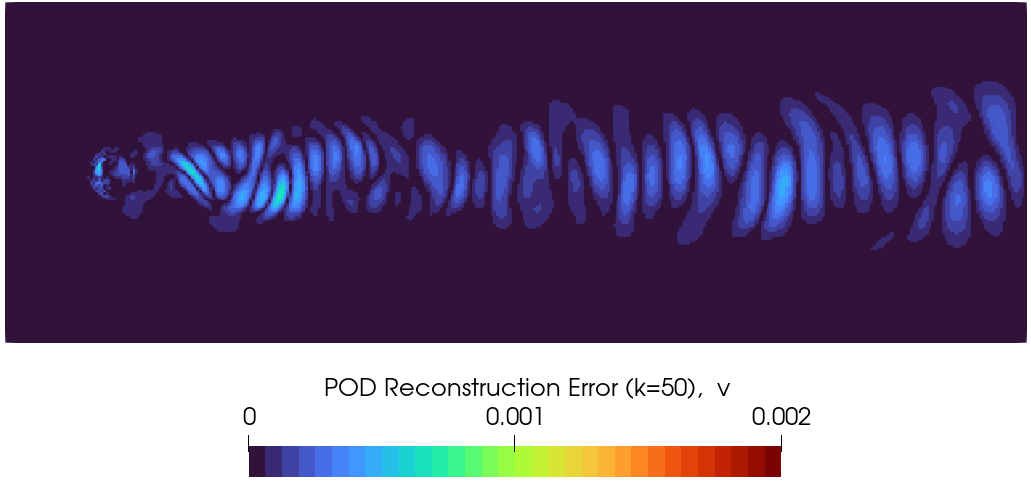}

  \caption{Reconstruction errors at $T = 5$ in $u$ and $v$ between POD and CAE ($k$ = 4) at $\bm{\mu^{*}} = [51, 142.2]$.}
  \label{fig:cylinder_recon}
  \end{figure}

\begin{figure}[!h]
  \centering

  \includegraphics[width=0.25\textwidth]{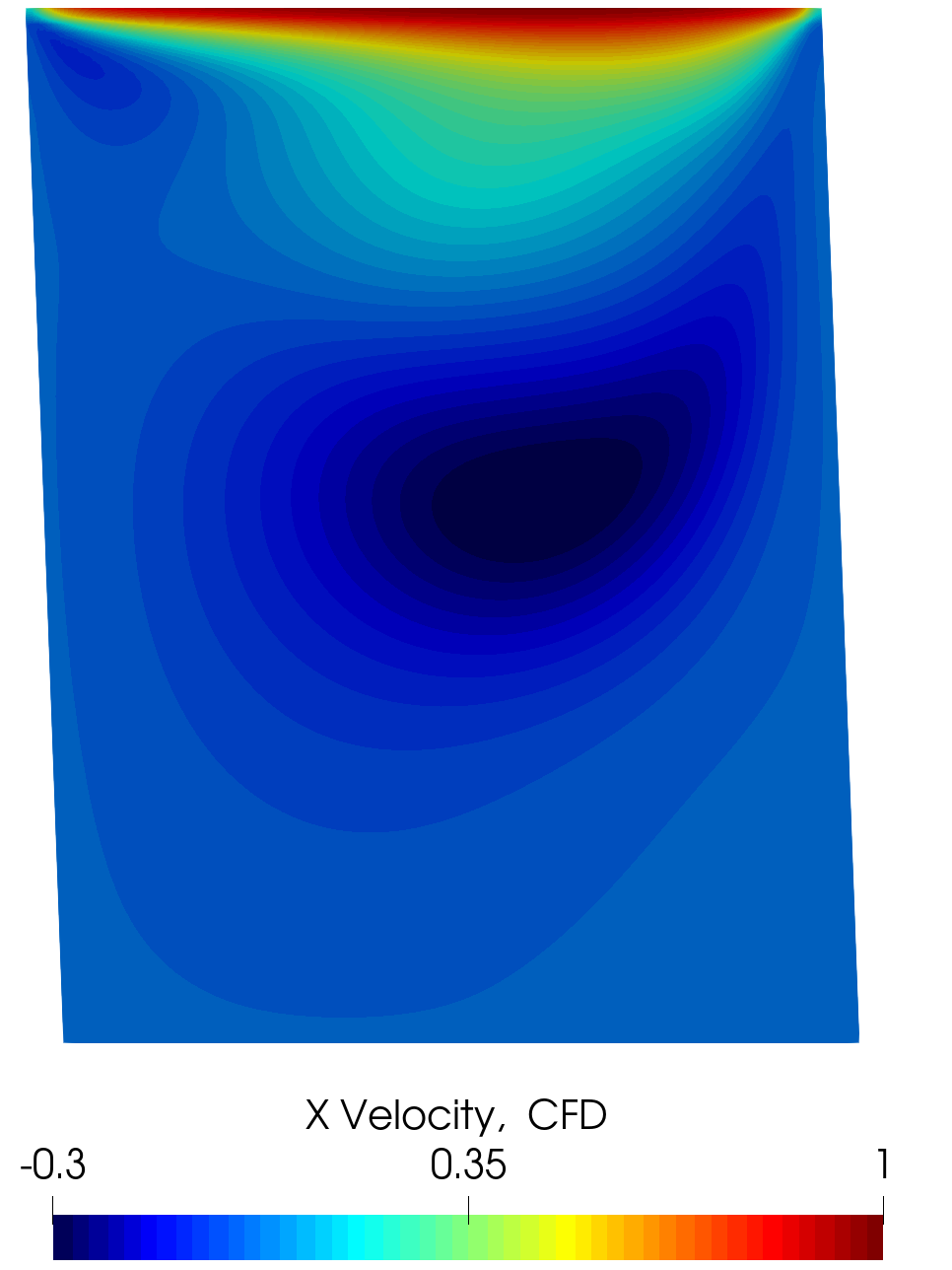}

  \includegraphics[width=0.25\textwidth]{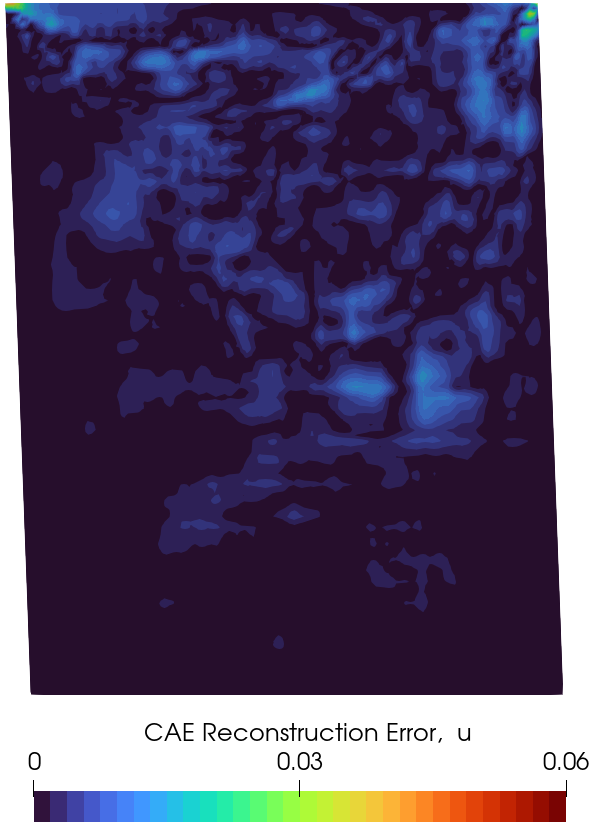}
  \includegraphics[width=0.25\textwidth]{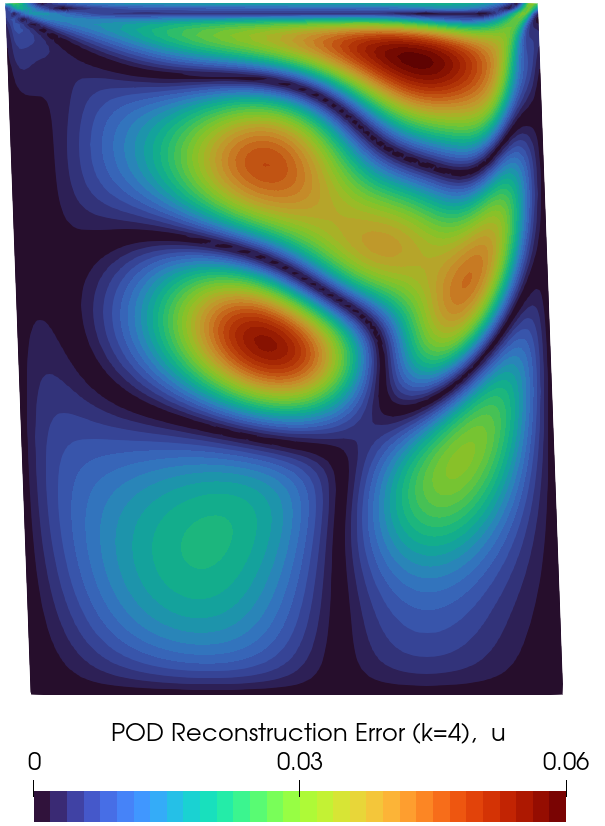}
  \includegraphics[width=0.25\textwidth]{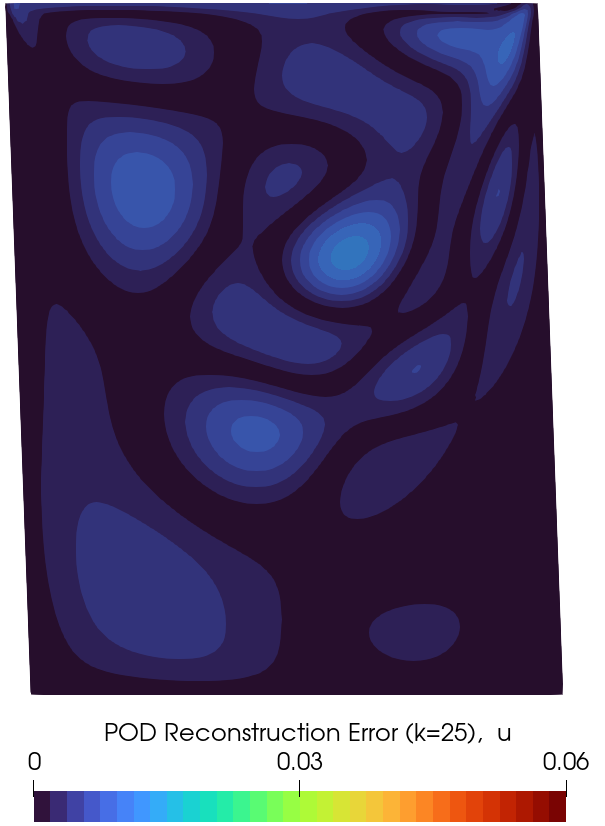}
  
  \includegraphics[width=0.25\textwidth]{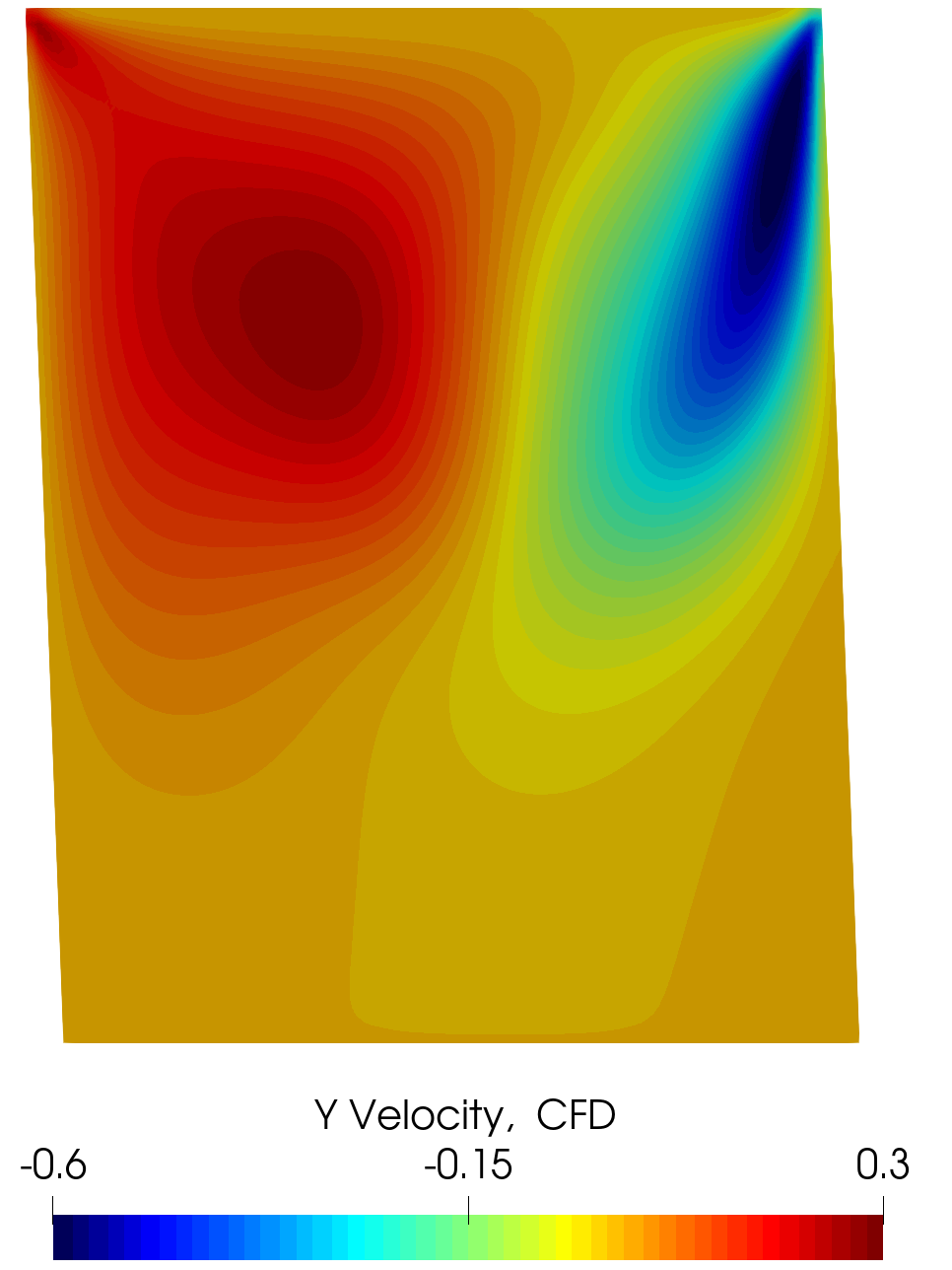}
  
  \includegraphics[width=0.25\textwidth]{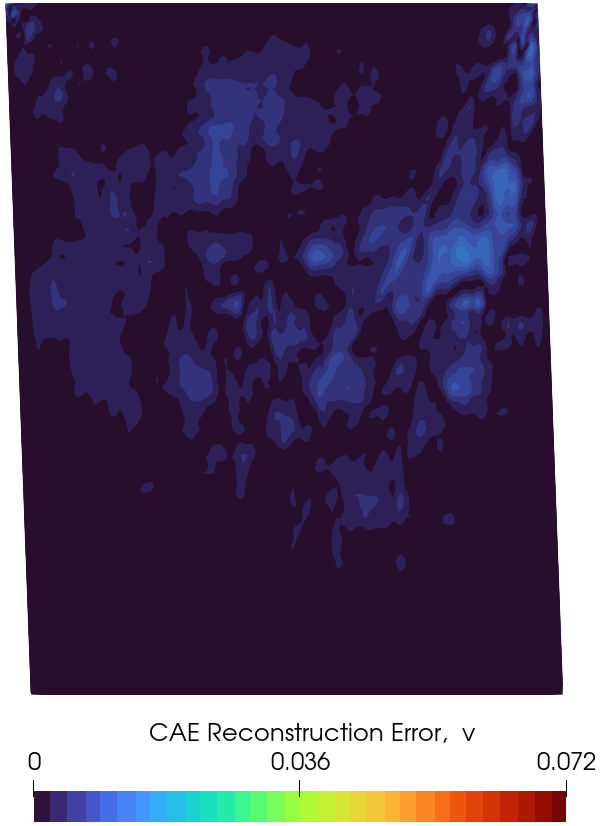}
  \includegraphics[width=0.25\textwidth]{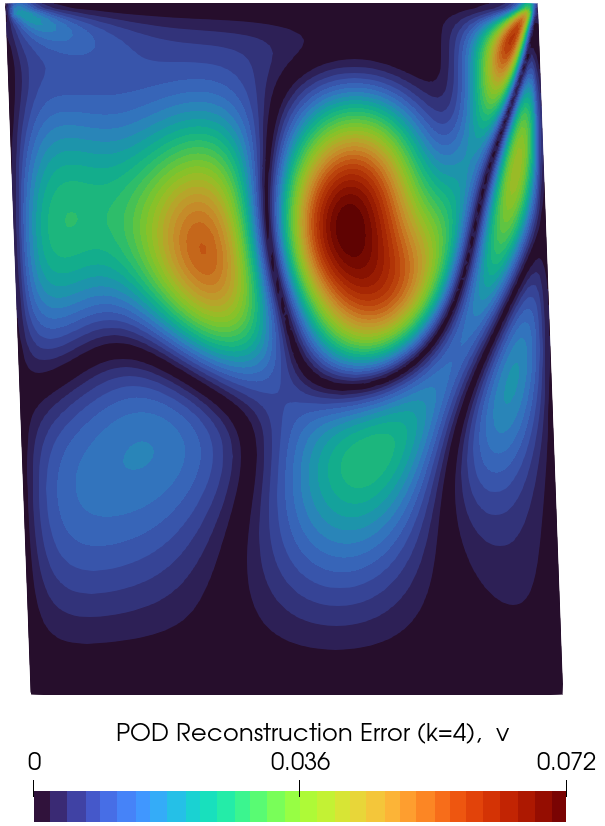}
  \includegraphics[width=0.25\textwidth]{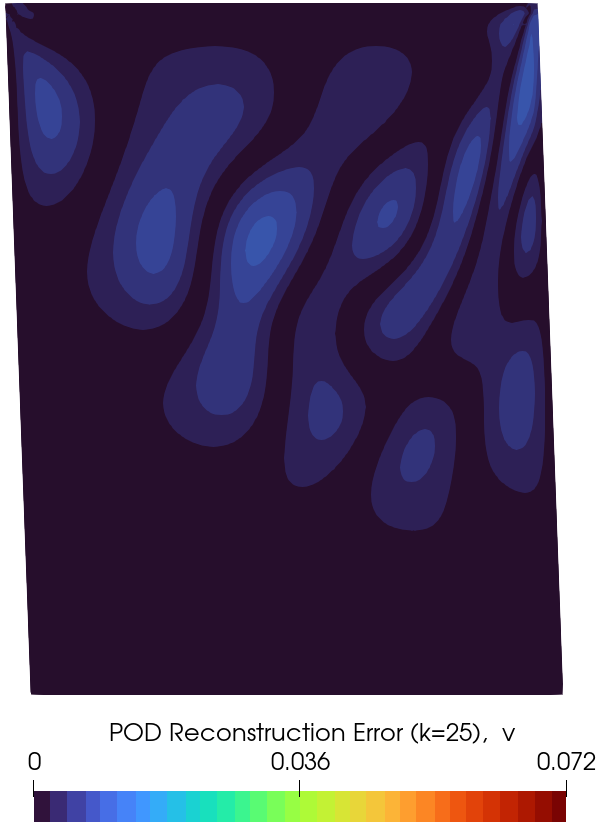}

  \caption{Reconstruction errors at $T = 5$ in $u$ and $v$ between POD and CAE ($k$ = 4) at $\bm{\mu^{*}} = [1.299, 1.689, -0.0367]$.}
  \label{fig:cavity_recon}
  \end{figure}

Reconstruction error comparisons between CAE and POD in $u$ and $v$ are given for both test cases. The same test points from the results section are used. Figure~\ref{fig:cavity_recon} shows the reconstruction errors between CAE ($k$ = 4) and POD at $k$ = 4 and $k$ = 25 for the lid-driven cavity case at $\bm{\mu^{*}} = [1.299, 1.689, -0.0367]$. For the same latent dimension, the CAE reconstruction errors are significantly smaller than those from POD throughout the domain in both $u$ and $v$. For $k = 25$, reconstruction errors from POD are similar to that of CAE. Figure~\ref{fig:cylinder_recon} shows reconstruction errors for the cylinder case at $\bm{\mu^{*}} = [51, 142.2]$. Again, it is shown that for the same latent dimension, the reconstruction error from a CAE is much smaller than from POD. At $k$ = 50, the error contours are comparable. Using POD would require the LSTM model to handle a very large multivariate time-series problem, which would degrade the accuracy of predictions.

\clearpage

\section{Convolutional autoencoder architectures}
\label{appendix:arch}
The convolutional autoencoder architectures for the lid-driven cavity and 2D cylinder cases are given in Tables~\ref{tab:cavity_cae} and~\ref{tab:cylinder_cae} respectively. Both architectures are similar and use convolutional layers in the encoder section to progressively reduce the number of pixels or grow the number of filters in subsequent layers. Convolutional layers are usually followed max-pooling layers that have batch normalization~\cite{ioffe2015batch} applied to them (referred to as pool-norm layers). Batch normalization normalizes the input of each layer over a mini-batch, reducing internal covariate shift, leading to more efficient gradient flow during backpropagation. Fully connected layers are present in both the encoder and decoder. Although their inclusion leads to a larger number of network parameters, reconstruction accuracies are significantly improved. The decoder consists of convolutional transpose layers that are used to progressively increase the number of pixels. The output layer uses a sigmoid activation function to constrain the range of outputs to [0,1]. The leaky ReLU activation function is used for convolutional and fully connected layers, and is given as,

\begin{equation}
  \phi(x) =
  \begin{cases}
  x,& \text{if } x\geq 0\\
  \alpha x,              & \text{if } x < 0
\end{cases}.
\end{equation}

A value of $\alpha$ = 0.25 is used for all leaky ReLU activation functions. The activation function is often used over the standard ReLU activation function ($\alpha$ = 0). For inputs less than 0, the ReLU activation function returns a gradient of zero, effectively rendering certain neurons inactive and inhibiting gradient flow, referred to as the dying ReLU problem.

\begin{table}[!h]
  \centering
  \begin{tabular}{@{}lllll@{}}
  \toprule
  {Layer}                          & {Number of Filters} & {Kernel Size} & {Activation Function} & {Size of Output} \\ \midrule
  Input                          &                   &             &                     & 128 $\times$ 128 $\times$ 2  \\ 
  Convolutional                  & 8                 & 3 $\times$ 3       & Leaky ReLU          & 128 $\times$ 128 $\times$ 8  \\ 
  Pool-Norm                      &                   & 2 $\times$ 2       &                     & 64 $\times$ 64 $\times$ 8    \\ 
  Convolutional                  & 16                & 3 $\times$ 3       & Leaky ReLU          & 64 $\times$ 64 $\times$ 16   \\ 
  Pool-Norm                      &                   & 2 $\times$ 2       &                     & 32 $\times$ 32 $\times$ 16   \\ 
  Convolutional                  & 32                & 3 $\times$ 3       & Leaky ReLU          & 32 $\times$ 32 $\times$ 32   \\ 
  Pool-Norm                      &                   & 2 $\times$ 2       &                     & 16 $\times$ 16 $\times$ 32   \\ 
  Convolutional                  & 64                & 3 $\times$ 3       & Leaky ReLU          & 16 $\times$ 16 $\times$ 64   \\ 
  Pool-Norm                      &                   & 2 $\times$ 2       &                     & 8 $\times$ 8 $\times$ 64     \\ 
  Reshape                        &                   &             &                     & 4096           \\ 
  Fully Connected                &                   &             & Leaky ReLU          & 128            \\ 
  Batch Norm                     &                   &             &                     & 128            \\ 
  Fully Connected (Latent Space) &                   &             &                     & 4           \\ 
  Fully Connected                &                   &             & Leaky ReLU          & 128            \\ 
  Batch Norm                     &                   &             &                     & 128            \\ 
  Fully Connected                &                   &             & Leaky ReLU          & 4096           \\ 
  Batch Norm                     &                   &             &                     & 4096           \\ 
  Reshape                        &                   &             &                     & 8 $\times$ 8 $\times$ 64     \\ 
  Convolutional Transpose        & 32                & 3 $\times$ 3       & Leaky ReLU          & 16 $\times$ 16 $\times$ 32   \\ 
  Batch Norm                     &                   &             &                     & 16 $\times$ 16 $\times$ 32   \\ 
  Convolutional Transpose        & 16                & 3 $\times$ 3       & Leaky ReLU          & 32 $\times$ 32 $\times$ 16   \\ 
  Batch Norm                     &                   &             &                     & 32 $\times$ 32 $\times$ 16   \\ 
  Convolutional Transpose        & 8                 & 3 $\times$ 3       & Leaky ReLU          & 64 $\times$ 64 $\times$ 8    \\ 
  Batch Norm                     &                   &             &                     & 64 $\times$ 64 $\times$ 8    \\ 
  Convolutional Transpose        & 2                 & 3 $\times$ 3       & Sigmoid          & 128 $\times$ 128 $\times$ 2  \\ \bottomrule
  
  \end{tabular}
  \caption{Convolutional autoencoder architecture for the lid-driven cavity case.}
  \label{tab:cavity_cae}
  \end{table}

  \begin{table}[!h]
    \centering
    \begin{tabular}{@{}lllll@{}}
    \toprule
    {Layer}                          & {Number of Filters} & {Kernel Size} & {Activation Function} & {Size of Output} \\ \midrule
    Input                          &                   &             &                     & 384 $\times$ 128 $\times$ 2  \\ 
    Convolutional                  & 8                 & 5 $\times$ 5       & Leaky ReLU          & 192 $\times$ 64 $\times$ 8   \\ 
    Batch Norm                     &                   &             &                     & 192 $\times$ 64 $\times$ 8   \\ 
    Convolutional                  & 16                & 5 $\times$ 5       & Leaky ReLU          & 96 $\times$ 32 $\times$ 16   \\ 
    Batch Norm                     &                   &             &                     & 96 $\times$ 32 $\times$ 16   \\ 
    Convolutional                  & 32                & 3 $\times$ 3       & Leaky ReLU          & 96 $\times$ 32 $\times$ 32   \\ 
    Pool-Norm                      &                   & 2 $\times$ 2       &                     & 48 $\times$ 16 $\times$ 32   \\ 
    Convolutional                  & 64                & 3 $\times$ 3       & Leaky ReLU          & 48 $\times$ 16 $\times$ 64   \\ 
    Pool-Norm                      &                   & 2 $\times$ 2       &                     & 24 $\times$ 8 $\times$ 64    \\ 
    Reshape                        &                   &             &                     & 12288          \\ 
    Fully Connected                &                   &             & Leaky ReLU          & 128            \\ 
    Batch Norm                     &                   &             &                     & 128            \\ 
    Fully Connected (Latent Space) &                   &             &                     & 10             \\ 
    Fully Connected                &                   &             & Leaky ReLU          & 128            \\ 
    Batch Norm                     &                   &             &                     & 128            \\ 
    Fully Connected                &                   &             & Leaky ReLU          & 12288          \\ 
    Batch Norm                     &                   &             &                     & 12288          \\ 
    Reshape                        &                   &             &                     & 24 $\times$ 8 $\times$ 64    \\ 
    Convolutional Transpose        & 32                & 3 $\times$ 3       & Leaky ReLU          & 48 $\times$ 16 $\times$ 32   \\ 
    Batch Norm                     &                   &             &                     & 48 $\times$ 16 $\times$ 32   \\ 
    Convolutional Transpose        & 16                & 3 $\times$ 3       & Leaky ReLU          & 96 $\times$ 32 $\times$ 16   \\ 
    Batch Norm                     &                   &             &                     & 96 $\times$ 32 $\times$ 16   \\ 
    Convolutional Transpose        & 8                 & 5 $\times$ 5       & Leaky ReLU          & 192 $\times$ 64 $\times$ 8   \\ 
    Batch Norm                     &                   &             &                     & 192 $\times$ 64 $\times$ 8   \\ 
    Convolutional Transpose        & 2                 & 5 $\times$ 5       & Sigmoid             & 384 $\times$ 128 $\times$ 2  \\ \bottomrule
    \end{tabular}
    \caption{Convolutional autoencoder architecture for the 2D cylinder case.}
    \label{tab:cylinder_cae}
    \end{table}

\end{appendices}

\clearpage
\bibliographystyle{unsrt}
\bibliography{references}

\end{document}